\pdfoutput=1 
\documentclass[twoside,english]{elsarticle}
\usepackage[LGR,T1]{fontenc}
\usepackage[latin9]{inputenc}
\usepackage{geometry}
\usepackage{array}
\usepackage{float}
\usepackage{multirow}
\usepackage{amsmath}
\usepackage{stackrel}
\usepackage{graphicx}
\usepackage{esint}

\makeatletter

\DeclareRobustCommand{\greektext}{%
  \fontencoding{LGR}\selectfont\def\encodingdefault{LGR}}
\DeclareRobustCommand{\textgreek}[1]{\leavevmode{\greektext #1}}

\providecommand{\tabularnewline}{\\}

\let\today\relax
\makeatletter
\def\ps@pprintTitle{%
    \let\@oddhead\@empty
    \let\@evenhead\@empty
    \def\@oddfoot{\footnotesize\itshape
         {} \hfill\today}%
    \let\@evenfoot\@oddfoot
    }
\makeatother

\usepackage{letltxmacro}
\LetLtxMacro{\originaleqref}{\eqref}
\renewcommand{\eqref}{Eq.~\originaleqref}

\makeatother

\usepackage{babel}
\begin{document}
\begin{frontmatter}
\title{Fickian Insights Using Probability Theory as Logic}
\author{Peter E. Price, Jr.}
\ead{learningtodry@gmail.com}
\address{LTD\&G LLC, 4652 Vincent Avenue South, Minneapolis, MN 55410}
\begin{abstract}
In \textit{Clearing Up Mysteries - The Original Goal} (Maximum Entropy
and Bayesian Methods: Cambridge, England, 1988. Springer, pp. 1\textendash 27),
Jaynes derived Fick's Law for a dilute binary solution from Bayes'
Theorem by considering, probabilistically, the motion of dilute solute
molecules. Modifying Jaynes' prior, changing the frame of reference,
and allowing for multicomponent systems, one can follow Jaynes' logic
to arrive at several expressions for the diffusion coefficient that
are widely used in application to solvent-polymer systems. These results,
however, do not generally satisfy required conditions over the full
concentration range. This limitation is resolved by considering the
joint motion of all components in the solution with the inclusion
of known physical constraints. Doing so, one arrives at a new set
of constitutive equations for binary and multicomponent diffusion
that include Darken's form as a limit and provide means to characterize
inter-species correlations, removing the need for the \textit{ad hoc}
modifications of the thermodynamic term or substitution of bulk compositions
with local compositions that have been proposed in the literature.
In binary systems, the Bayesian expression for the mutual diffusion
coefficient provides explicit limits on non-ideal mutual diffusion
behavior, given component self-diffusion coefficients, that can be
attributed to inter-species correlation versus stronger intra- and
inter-species clustering. Applications of the model to published data
for a set of binary systems are presented, along with pointers to
further directions for research.
\end{abstract}
\end{frontmatter}

\section{Introduction}

Fickian representations of diffusion describe component fluxes relative
to a reference velocity as the products of mutual diffusion coefficients
and gradients in some measure of concentration. For one dimensional
molar diffusion in a binary system relative to the volume average
velocity, Fick\textquoteright s law has the form:

\begin{equation}
J_{i}^{V}=-D\frac{dC_{i}}{dx}\label{eq:Fick_1}
\end{equation}

Based upon consideration of the correspondence between the Maxwell-Stefan
(MS) and Fickian representations of diffusion, the Fickian mutual
diffusion coefficients can be separated into the product of terms
characterizing the frictional resistance to motion between diffusing
entities, \DJ , and a thermodynamic correction factor, \textgreek{\textGamma}.
The thermodynamic factor is derived from the understanding that gradients
in chemical potential, not concentration, are what drive diffusive
fluxes \citep{taylor1993multicomponent,krishna2015uphilldiffusion}: 

\begin{equation}
D=\mkern2mu\rule[0.75ex]{0.75ex}{0.06ex}\mkern-8mu D\Gamma=\mkern2mu\rule[0.75ex]{0.75ex}{0.06ex}\mkern-8mu D\left(1+\frac{dln\left(\gamma_{1}\right)}{dln\left(x_{1}\right)}\right)\label{eq:Fick_2}
\end{equation}
The literature contains a variety of approaches to constructing the
frictional part from the constituent self-diffusion coefficients,
many starting from Darken's \citep{darken1948diffusion} binary form: 

\begin{equation}
\mkern2mu\rule[0.75ex]{0.75ex}{0.06ex}\mkern-8mu D=x_{1}D_{2}+x_{2}D_{1}\label{eq:Darken}
\end{equation}
where $x_{i}$ are component mole fractions, and $D_{i}$ are component
self-diffusion coefficients.

Another line of expressions relating mutual and self-diffusion coefficients
have followed from Vignes' \citep{vignes1966diffusion} empirical
proposal: 

\begin{equation}
\mkern2mu\rule[0.75ex]{0.75ex}{0.06ex}\mkern-8mu D=D_{1}^{x_{2}}D_{2}^{x_{1}}\label{eq:Vignes}
\end{equation}

The self-diffusion coefficients characterize the mean square displacements
of species in a given time \citep{einstein1956investigations} in
the absence of chemical potential gradients, and thus characterize
the frictional resistance to motion of the individual components in
a solution. Since self-diffusion coefficients are relatively easy
to measure \citep{moggridge2012prediction,zhu2015alocal}, there is
strong motivation to develop a robust theory relating mutual diffusion
coefficients to constituent self-diffusion coefficients. Though tremendous
progress has been made \citep{alabi-babalola2024rationalizing}, elements
of that progress remain linked to model forms that have found success
through empirical modifications. 

For example, recognition that there are often associations between
components, both like and unlike, that can lead to cluster formation
has led to modifications of both the frictional and the thermodynamic
components in the mutual diffusion coefficient expressions. Several
of these \citep{carman1967selfdiffusion,moggridge2012prediction}
take the general form:

\begin{equation}
D=(x_{1}D_{2}+x_{2}fD_{1})\Gamma^{\alpha}\label{eq:Moggridge}
\end{equation}
where $f$ is a function \citep{carman1967selfdiffusion} or constant
\citep{moggridge2012prediction} reflecting like-molecule polymerized
clusters and $\alpha$ (<1) is a constant adapted from considerations
of diffusion near consolute points \citep{dagostino2011prediction,moggridge2012prediction2}
to dampen the thermodynamic term. For modeling systems that include
$f$ across the full concentration range, concentration dependence
in $f$ is required for the model to reach the appropriate pure component
limits.

Alternatively, several authors have shown improved agreement with
experimental data by replacing the mole fractions in \DJ{} with local
mole fractions \citep{li2001amutualdiffusioncoefficient,zhou2013localcomposition,zhu2015alocal}
to account for molecular associations. However, as \citet{jianmin2008phenomenological2}
note in regard to substitution of local for bulk compositions in binary
models, multicomponent models based on local compositions must satisfy
material balance constraints, Gibbs-Duhem consistency, and the Onsager
reciprocal relations (ORR) to be broadly applicable. Substitution
of local compositions for bulk values can lead to violations of these
conditions.

Compilations of proposed expressions for binary mutual diffusion coefficients
in terms of component self-diffusion coefficients have been presented
by \citet{hsu1998correlation}, \citet{guevara-carrion2016mutualdiffusion},
\citet{obukhovsky2017nonlinear}, and most recently by \citet{alabi-babalola2024rationalizing}.
All of these models are based on physical or empirical considerations
of diffusion processes. By shifting perspective to an information
theoretic viewpoint that incorporates prior physical knowledge, a
new set of expressions for the relationship between self- and mutual
diffusion coefficients can be derived for both binary and multicomponent
systems. This approach follows and extends Jaynes\textquoteright{}
\citep{jaynes1989clearing} use of probability theory as logic to
derive Fick\textquoteright s law for a binary system from Bayes\textquoteright{}
Theorem. The resultant expressions provide new insights into existing
theories and, since they satisfy required constraints, may well be
useful in their own right. 

Because this approach is very different from existing lines of development,
we will limit our review of those existing lines to the above discussion
and welcome the reader to consult the references. Below, we review
Jaynes\textquoteright{} derivation in more detail and show how several
existing theories in common application for solvent-polymer systems
\citep{zielinski1999practical,alsoy1999modeling} follow from Jaynes\textquoteright{}
approach if we allow multiple components, alternative frames of reference,
and, most importantly, if we recognize and include our prior information
that chemical potentials drive the motion. The unifying aspect of
these derivations is that they consider only information about a single
species. Failure to put constraints on the joint movements in multicomponent
systems results in expressions that satisfy the ORR only under special
conditions \citep{price2003multicomponent}. The present approach
sheds new light on the information content of these existing theories
and points the way to further development. By considering the joint
information in a binary system, we extend Jaynes\textquoteright{}
binary results to a new expression for the mutual diffusion coefficient
that includes the self-diffusion coefficients of both components and
a function that accounts for correlated motion of the components.
To understand the range of possibilities for this model, we compare
its results relative to several other theories for some nonphysical
parameter sets, and then examine its performance relative to a sample
of published experimental data. Finally, we extend the joint information
derivation to multicomponent systems and provide a comparison for
a ternary system to the multicomponent form of Darken\textquoteright s
equation \citep{price2003multicomponent} derived from Bearman's \citep{bearman1961onthe}
friction-based approach.

Hopefully the reader, like Jaynes, will find excitement and opportunity
in failure, because the Bayesian model, while showing insightful success
for some systems, shows spectacular failure in others. In many cases,
the failure may be due to our own failure to account for uncertainty
in the underlying characterization of a given system's self-diffusion
and thermodynamic behavior. In other cases, the failure signal is
robust and attributable to omission of relevant physical information
in the theory, appearing when strong clustering is expected. In that
sense, the model delineates behavior that can be explained by increased
friction, and hence correlation structure, in the self-diffusion motions
from that which requires more complete accounting of cluster formation.

\section{Probability and Bayes' Theorem}

Probability is a concept that has a variety of interpretations \citep{howie2002interpreting},
and its foundations have been rife with controversies, both philosophical
and mathematical. Because Jaynes' results are extended here, his framework
for probability \citep{jaynes2003probability} will be followed without
further justification. To wit, the probability $P\left(x\mid I\right)$
of a proposition $x$ is taken to represent degree of belief in $x$
given one's specified or assumed state of knowledge about the problem
at hand, $I$, on a scale of real numbers spanning from $P\left(x\mid I\right)=0$
when $x$ is known to be false to $P\left(x\mid I\right)=1$ when
$x$ is known to be true. Elements to the right of the vertical line
in this notation are taken to be true. The probabilites of all mutually
exclusive propositions must sum to 1. Calculations with probabilities
rely on the product rule for joint propositions:

\begin{equation}
P\left(x,y\mid I\right)=P\left(x\mid I\right)P\left(y\mid x,I\right)=P\left(y\mid I\right)P\left(x\mid y,I\right)
\end{equation}
And the sum rule for a proposition $x$ and its negation $\overline{x}$:

\begin{equation}
P\left(x\mid I\right)+P\left(\overline{x}\mid I\right)=1
\end{equation}

Bayes' Theorem is a logical outcome of these rules, and for many practical
applications in the physical sciences including the application described
below, takes the form:

\begin{equation}
P\left(h\mid d,I\right)=\frac{P\left(h\mid I\right)P\left(d\mid h,I\right)}{P\left(d\mid I\right)}\label{eq:Bayes-1}
\end{equation}
where $h$ is represents a hypothesis, and $d$ represents the data.
The term on the left is the posterior probability of hypothesis $h$
given data $d$ and other information in the problem formulation,
$I$. The first term in the numerator on the right is the prior probability
for hypothesis $h$. Priors incorporate information regarding $h$
available or assumed before learning from data. If one begins an inference
from a different state of information, one's prior may change, leading
to a different posterior. The second term in the numerator on the
right is the likelihood of the data $d$ given hypothesis $h$. The
denominator on the right is referred to as the evidence for data $d$
and serves as a normalizing constant for the posterior.

\section{Inference on Individual Component Motions}

\subsection{Jaynes\textquoteright{} Derivation of Fick\textquoteright s Law for
a Binary System from Bayes\textquoteright{} Theorem}

Jaynes \citep{jaynes1989clearing} considered a solution of sugar
in water \textquotedblleft so dilute that each sugar molecule interacts
constantly with the surrounding water, but almost never encounters
another sugar molecule.\textquotedblright{} Given a nonuniform concentration
profile in one dimension, he wrote the species conservation law for
sugar and then, without stating so explicitly, expressed the flux
as Fick\textquoteright s Law relative to the volume average velocity:

\begin{equation}
J_{1}^{V}=-D\frac{dC_{1}}{dx}
\end{equation}
where $C_{1}$ is the sugar concentration, and $J_{1}^{V}$ is the
diffusive flux of sugar, D is the mutual diffusion coefficient. The
form of Jaynes\textquoteright{} conservation equation implies an assumption
that the volume average velocity was zero. Jaynes also made no stated
consideration of hydration, an issue of importance to the sugar-water
system he chose for illustrative purposes and related to various theories
of cluster diffusion mentioned above.

Fick (republished \citep{fick1995onliquid}) proposed a molecular
diffusive flux relation based upon the gradient of concentration.
He did so on phenomenological grounds, by analogy to Fourier\textquoteright s
and Ohm\textquoteright s Laws for the diffusion of thermal energy
and electricity, respectively. Jaynes sought a more fundamental basis
for this expression and began by writing a centered difference expression
for the velocity, $v$, of a molecule of sugar presently located at
position $x(t)$:

\begin{equation}
v=\frac{x\left(t+\tau\right)-x\left(t-\tau\right)}{2\tau}=\frac{y-z}{2\tau}\label{eq:velocity}
\end{equation}

The averaging time, $\tau$, is assumed long with respect to the time
scale of the rapid thermal fluctuations of the molecules in solution.
Next, noting that movement of the sugar molecule in the dilute solution
is the result of many small collisions, primarily between the sugar
molecule and surrounding water, Jaynes invoked the Central Limit Theorem
to give a Gaussian probability distribution for the future location
of a molecule, $y=x(t+\tau)$, given its present position, $x(t)$:

\begin{equation}
P\left(y\mid x,\tau,I\right)=A_{1}exp\left[-\frac{\left(y-x\right){}^{2}}{2\left(\sigma\left(\tau\right)\right)^{2}}\right]\label{eq:likelihood}
\end{equation}
where \textquotedblleft $I$ stands for the general prior information
stated or implied in our formulation of the problem\textquotedblright{}
\citep{jaynes1989clearing}, and $A_{1}$ is a normalizing constant.
The \textquotedblleft spreading function,\textquotedblright{} $\left(\sigma\left(\tau\right)\right)^{2}$,
is the expected square of the molecule\textquoteright s displacement
in time $\tau$, $\left(\sigma\left(\tau\right)\right)^{2}=\left(\delta x\right){}^{2}$.
The probability distribution in \eqref{eq:likelihood} is symmetric
about the present position of the molecule, and so the expected and
most probable values of y are just the current position. Jaynes noted
that the time-reversal invariance of the equations of motion suggests
that the same form should hold for the past position of the molecule,
$z=x\left(t-\tau\right)$. This implies that the expected velocity
of the molecule, \eqref{eq:velocity}, should be zero. The fact that
gradients in such a dilute system do level out with time indicates
that average velocities are not zero, presenting an apparent paradox.
Jaynes addressed this apparent paradox by noting that, \textquotedblleft The
equations of motion are symmetric in past and future, but our information
about the particles is not\textquotedblright{} \citep{jaynes1989clearing}.
Our information about the past concentration profile is what breaks
the symmetry of the predicted movement. He then wrote Bayes\textquoteright{}
Theorem for the probability distribution of past location, $z$, of
the sugar molecule given its present position, $x$, the time step,
$\tau$, and the rest of our prior information, $I$:

\begin{equation}
P\left(z\mid x,\tau,I\right)=A_{2}P\left(z\mid\tau,I\right)P\left(x\mid z,\tau,I\right)\label{eq:Bayes_sugar}
\end{equation}
where, again, $A_{2}$ is a normalizing constant, the inverse of the
evidence, $P\left(x\mid\tau,I\right)$. Replacing $y$ with $x$ and
$x$ with $z$, \eqref{eq:likelihood} also represents the probability
distribution for the present molecule position, given its past position,
taken over the class of all motions allowed by the stated or implied
molecular dynamics, $P\left(x\mid z,\tau,I\right)$. It is our prior
knowledge of the concentration field that restricts the possible motions
within this class. Thus, Jaynes stated that the prior probability
of the molecule\textquoteright s position, $P\left(z\mid\tau,I\right)$,
is \textquotedblleft clearly proportional to $C_{1}\left(z\right)$.\textquotedblright{}
Inserting these equations into \eqref{eq:Bayes_sugar} and taking
the logarithm gives:

\begin{equation}
ln\left(P\left(z\mid x,\tau,I\right)\right)=ln\left(A_{2}\right)+ln\left(C_{1}\left(z\right)\right)-\frac{\left(x-z\right){}^{2}}{2\left(\sigma\left(\tau\right)\right)^{2}}
\end{equation}

The most probable past position of the molecule now at $x$ is found
by differentiating with respect to $z$ and setting the result equal
to zero. Solving the resulting equation for the most probable past
position of the molecule gives: 

\begin{equation}
\hat{z}=x+\left(\sigma\left(\tau\right)\right)^{2}\frac{dln\left(C_{1}\left(z\right)\right)}{dz}=x+\left(\delta x\right){}^{2}\frac{dln\left(C_{1}\right)}{dz}
\end{equation}

Jaynes appears to have assumed that the gradient term, evaluated at
time $\tau$ in the past and location $\hat{z}$, was sufficiently
close to its present value at $x$ to substitute this result into
\eqref{eq:velocity} to write the most probable drift velocity as:

\begin{equation}
\hat{v}=-\frac{\left(\delta x\right){}^{2}}{2\tau}\frac{dln\left(C_{1}\right)}{dx}
\end{equation}
This leads to the flux expression:

\begin{equation}
J_{1}^{V}=C_{1}\hat{v}=-\frac{\left(\delta x\right){}^{2}}{2\tau}\frac{dC_{1}}{dx}=-D_{1}\frac{dC_{1}}{dx}=-D\frac{dC_{1}}{dx}\label{eq:Bayesian=000020flux}
\end{equation}

The interested reader will find much more to contemplate in Jaynes\textquoteright{}
\citep{jaynes1989clearing} discussion, but we note that the mutual
diffusion coefficient, $D$, is the self-diffusion coefficient of
the sugar in the dilute solution, $D_{1}$, defined in terms of a
sugar molecule\textquoteright s expected square displacement in time
. This is Einstein\textquoteright s expression for the (tracer or
self-) diffusion coefficient in Brownian motion (reprinted \citet{einstein1956investigations}).
While one may take exception to any of Jaynes\textquoteright{} assumptions,
since Bayes\textquoteright{} Theorem is a consequence of the product
and sum rules of probability, Fick\textquoteright s Law and Einstein\textquoteright s
expression for $D$ are consistent outcomes of applying probability
theory as logic given the stated assumptions. \eqref{eq:Bayesian=000020flux}
is also consistent with the fundamental result that in the dilute
limit, the mutual diffusion coefficient in a binary system approaches
the self-diffusion coefficient of the dilute component.

\subsection{Fick's Law from the Mean}

Rather than considering only the most probable past position of a
molecule, we can derive the expected (mean) location by linearizing
the concentration-based prior about the current position, which we
take as $x=0$:

\begin{equation}
P\left(z\mid x=0,\tau,I\right)=A_{3}\left(C_{1,x=0}+\frac{dC_{1}}{dx}_{x=0}z\right)exp\left[-\frac{z^{2}}{2\left(\sigma\left(\tau\right)\right)^{2}}\right]
\end{equation}

$A_{3}$ is a constant. Again, the gradient is assumed constant on
the time scale of interest. If we treat the domain as infinite, then
far from the current position, $x=0$, any non-zero gradient in the
linearized concentration prior will lead to invalid negative probabilities.
However, we proceed on the assumption that the likelihood function
decays much faster than the gradient term and renders such errors
negligible for present purposes. The normalizing factor is given by
the integral:

\begin{equation}
P\left(x=0\mid\tau,I\right)=\intop_{-\infty}^{\infty}A_{3}\left(C_{1,x=0}+\frac{dC_{1}}{dx}_{x=0}z\right)exp\left[-\frac{z^{2}}{4\tau D_{1}}\right]dz=\sqrt{4\pi\tau D_{1}}A_{3}C_{1,x=0}
\end{equation}
And the expected (mean) past position of the molecule now at $x=0$
is:

\begin{equation}
\left\langle z\right\rangle =\intop_{-\infty}^{\infty}\frac{z\left(C_{1,x=0}+\frac{dC_{1}}{dx}_{x=0}z\right)exp\left[-\frac{z^{2}}{4\tau D_{1}}\right]}{\sqrt{4\pi\tau D_{1}}C_{1,x=0}}dz=\frac{2\tau D_{1}}{C_{1,x=0}}\frac{dC_{1}}{dx}_{x=0}
\end{equation}
This leads, once again, to Fick's Law:

\begin{equation}
J_{1}^{V}=C_{1}\left\langle v\right\rangle =C_{1}\frac{-\left\langle z\right\rangle }{2\tau}=-D_{1}\frac{dC_{1}}{dx}_{x=0}=-D\frac{dC_{1}}{dx}
\end{equation}

\subsection{Multiple Components, Alternative Reference Frames, and a Potential
Prior}

Jaynes\textquoteright{} reasoning is based on consideration of a single
molecule, but the resultant expression is commonly applied in a continuum
context. In the spirit of investigation, we proceed allowing the continuum
concept of species coexistence in space while continuing to construct
expectations about individual component motions. Under such an assumption,
the logic of Jaynes\textquoteright{} derivation does not change if
there are multiple dilute components. Nor does it change if the molecular
motions occur relative to a moving frame of reference. Finally, diffusion
does not have to occur down a concentration gradient \citep{krishna2015uphilldiffusion}.
Fundamentally, diffusive motions occur from regions of higher chemical
potential to regions of lower chemical potential. This is prior information
that can be readily incorporated into Bayes' Theorem by replacing
Jaynes\textquoteright{} concentration-based past-position prior with
one based on component activity.

In a reference frame moving with velocity $v^{ref}$, the mapping
between the moving coordinate, $x$, and the laboratory coordinate,
$x^{lab}$, is: 

\begin{equation}
x^{lab}=x+v^{ref}t
\end{equation}

The position of a molecule of component $i$, presently at the origin
of the moving frame ($x_{i}=0$ at $t=0$), after time $t=\tau$ is
then: 

\begin{equation}
x_{i}\left(\tau\right)=\left(v_{i}-v^{ref}\right)\tau
\end{equation}
where $v_{i}$ is the particle velocity in the laboratory frame.

With respect to the moving reference frame, the species velocity,
centered at the present position and time ($x_{i}=0$ and $t=0$),
retains a form similar to \eqref{eq:velocity}: 

\begin{equation}
v_{i}-v^{ref}=\frac{x_{i}\left(\tau\right)-x_{i}\left(-\tau\right)}{2\tau}=\frac{y_{i}-z_{i}}{2\tau}=\frac{-z_{i}}{2\tau}\label{eq:moving=000020frame=000020velocity}
\end{equation}
where $y_{i}$ and $z_{i}$ are the future and past particle positions
of component $i$ in the moving frame of reference. If the reference
velocity represents some bulk translation of the components in such
a way that the molecular collisions can be considered as superimposed
on the motion described by the reference velocity, then \eqref{eq:likelihood}
will continue to hold for the future position of a molecule of component
$i$ in the moving reference frame. Under such conditions, the expected
future position of the molecule of $i$, $y_{i}$, is just its present
position, resulting in the last expression for the species velocity
given in \eqref{eq:moving=000020frame=000020velocity}.

The results that follow can also be derived by considering the expected
behavior using a linearization of the component activity about the
current position, analogous to the results in Section 3.2.

\subsubsection{Zielinski and Hanley\textquoteright s Model}

Consider, for example, mutual diffusion of a molecule of species $i$
relative to the mass average velocity, $v^{ref}=v^{m}$, of the system.
Recognizing that a molecule at some current position is more likely
to have come from a location of higher chemical potential in time
past, and letting species activity serve as a measure of that potential,
we replace Jaynes\textquoteright{} concentration-based prior with
a prior probability for the past position of that molecule proportional
to its activity, $a_{i}$. It is straightforward to follow Jaynes\textquoteright{}
derivation to arrive at the most probable past position: 

\begin{equation}
\hat{z_{i}}=x_{i}+(\sigma_{i}(\tau))^{2}\frac{dln\left(a_{i}\left(z\right)\right)}{dz}=x_{i}+\left(\delta x_{i}\right){}^{2}\frac{dln\left(a_{i}\left(z\right)\right)}{dz}
\end{equation}

Following the earlier results, the most probable species velocity
relative to the reference velocity becomes: 

\begin{equation}
\hat{v}_{i}-v^{m}=-\frac{\left(\delta x_{i}\right){}^{2}}{2\tau}\frac{dln\left(a_{i}\right)}{\partial x}
\end{equation}

The corresponding mass diffusion flux relative to the mass average
velocity is: 

\begin{equation}
j_{i}=\rho_{i}\left(\hat{v}_{i}-v^{m}\right)=-D_{i}\rho_{i}\frac{dln\left(a_{i}\right)}{dx}\label{eq:Zielinski}
\end{equation}
where $D_{i}$ is the component $i$ self-diffusion coefficient. For
multicomponent systems, the gradient term can be expanded in the component
concentrations using the chain rule. We can map the corresponding
species mass flux to the volume average reference frame \citep{haase1990thermodynamics2}
using the relation: 

\begin{equation}
j_{i}^{V}=j_{i}-\rho_{i}\stackrel[k=1]{N}{\sum}\hat{V}_{k}j_{k}=\left(1-\rho_{i}\left(\hat{V}_{i}-\hat{V}_{N}\right)\right)j_{i}-\rho_{i}\stackrel[k=1]{N-1}{\sum}\left(\hat{V}_{k}-\hat{V}_{N}\right)j_{k}
\end{equation}
where the $\hat{V}_{k}$ are the species partial specific volumes.
Substituting \eqref{eq:Zielinski} into this expression gives the
diffusion model presented for ternary systems by \citet{zielinski1999practical}.
The same model can be developed in the Bearman \citep{bearman1961onthe}
friction factor framework by assuming that the friction factors are
inversely proportional to the component molecular weights.

The literature contains some discussion of the Gibbs-Duhem consistency
of this model and its importance in satisfying material balance constraints
\citep{zielinski1999practical,nauman2001anengineering,price2003multicomponent}.
A related and detailed examination of restrictions on friction factors
imposed by both the entropy inequality and the Gibbs-Duhem equation
was presented by Vrentas and Vrentas \citep{vrentas2007restrictions}.
For the mass average reference frame, the component mutual diffusion
fluxes are constrained to satisfy the summation rule:

\begin{equation}
\stackrel[k=1]{N}{\sum}j_{k}=\stackrel[k=1]{N}{\sum}D_{k}\rho_{k}\frac{\partial ln\left(a_{k}\right)}{\partial x}=0
\end{equation}

and the isobaric, isothermal Gibbs-Duhem equation is:

\begin{equation}
\stackrel[k=1]{N}{\sum}\frac{\rho_{k}}{M_{k}}\frac{\partial\mu_{k}}{\partial x}=\stackrel[k=1]{N}{\sum}\frac{\rho_{k}}{M_{k}}\frac{\partial ln\left(a_{k}\right)}{\partial x}=0
\end{equation}

One can see by inspection that Gibbs-Duhem consistency requires that
all of the component self-diffusion coefficients, $D_{i}$, be inversely
proportional to the molecular weights:

\begin{equation}
D_{i}\propto\frac{1}{M_{i}}
\end{equation}

Such self-diffusion behavior is not typical. Our inferential derivation
of this mutual diffusion model, however, made no use of the Gibbs-Duhem
relationship. Because our inference is optimal in the Bayesian sense
given the assumptions described above, the present derivation may
help explain why Zielinski and Hanley\textquoteright s model performs
so well up to moderate solvent concentrations even in instances where
the self-diffusion coefficients do not satisfy the Gibbs-Duhem consistency
constraints. Likewise, the fact that we did not use any information
about the other species in the system of interest, and thus imposed
no a priori constraints on the individual species fluxes, explains
why failures of the theory occur when the proportionality constraint
is not met.

\subsubsection{Alsoy and Duda\textquoteright s Model}

It is also a straightforward exercise to follow this same logic using
the volume average velocity as the reference velocity. This leads
to the following expression for the mass diffusion flux of component
i: 

\begin{equation}
j_{i}^{V}=\rho_{i}\left(\hat{v}_{i}-v^{V}\right)=-D_{i}\rho_{i}\frac{dln\left(a_{i}\right)}{dx}
\end{equation}

This model was shown for ternary systems as Case 4 in the work of
\citet{alsoy1999modeling} and is the model derived using Bearman's
\citep{bearman1961onthe} friction-factor approach by assuming that
the friction factors are inversely proportional to the component partial
molar specific volumes. Consideration of the Gibbs-Duhem consistency
of this model also requires that the self-diffusion coefficients,
$D_{i}$, be inversely proportional to the component specific volumes.

\begin{equation}
D_{i}\propto\frac{1}{\hat{V}_{i}}
\end{equation}

Again, however, the model often performs quite well even when the
self-diffusion coefficients do not satisfy this constraint, suggesting
that its inferential basis helps explain why it is more broadly applicable
than might be expected from its mechanical foundations.

\section{Joint Inference on All Component Motions}

The problems that concern us in diffusion theory involve systems with
more than one component, and yet the derivations above used only information
about single components. By allowing the concept of species superposition,
we can now consider the problem of mutual diffusion as one of inference
on the joint probability distribution of past positions for molecules
of all species that are presently at the position of interest.

\subsection{Binary Systems}

Let us consider the situation in a binary system where molecules of
components 1 and 2 are presently centered at the origin ($x_{i}=0$
and $t=0$) of a frame of reference moving with velocity $v^{ref}$.
As above, if the reference velocity represents some bulk translation
of the components in such a way that the molecular collisions can
be considered as superimposed on the motion described by the reference
velocity, then \eqref{eq:likelihood} will continue to hold for the
future position of a molecule of component $i$ in the moving reference
frame. Jaynes' derivation considered a very dilute solution. In situations
where the components are not dilute but the gradients are not extreme,
the neighborhood of any molecule of interest on the relevant scales
of time and space can be assumed constant, collisions between like
and unlike molecules locally symmetric, and thus the Central Limit
Theorem still applicable. Let us further assume that inter-molecular
forces may result in associations between molecules of different species,
and thus in correlations between the local movements of those molecules.
Such associations form the conceptual foundation of local composition
theories, e. g., \citet{renon1968localcompositions}. Assuming probabilities
of prior positions are proportional to the species\textquoteright{}
activities, then the joint posterior distribution for the position
of the molecules of species 1 and 2, presently at the origin of the
moving reference frame, at time $\tau$ in the past is:

\begin{equation}
\begin{array}{c}
P(z_{1},z_{2}\mid x_{1}=0,x_{2}=0,\tau,I)=A_{4}\left(a_{1}(z_{1})\right)\left(a_{2}(z_{2})\right)exp\left[-\frac{\boldsymbol{z}^{T}\boldsymbol{\varSigma}^{-1}\boldsymbol{z}}{2}\right]\\
\boldsymbol{z}=\left[\begin{array}{c}
z_{1}\\
z_{2}
\end{array}\right],\boldsymbol{\varSigma}=\left[\begin{array}{cc}
\left(\sigma_{1}\left(\tau\right)\right){}^{2} & r_{12}\sigma_{1}\left(\tau\right)\sigma_{2}\left(\tau\right)\\
r_{12}\sigma_{1}\left(\tau\right)\sigma_{2}\left(\tau\right) & \left(\sigma_{2}\left(\tau\right)\right){}^{2}
\end{array}\right]
\end{array}\label{eq:Bayes=000020binary=000020joint}
\end{equation}
where $A_{4}$ is a normalizing constant, and $r_{12}$ is the correlation
coefficient for the motions of the two components.

Before proceeding, we will invoke and generalize our earlier connection
between the spreading functions, $\left(\sigma_{1}\left(\tau\right)\right){}^{2}$,
and the component self-diffusion coefficients, $D_{i}$. That is,
we assume that the spreading functions are locally constant for the
time and space scales of interest and that a more complete description
of the local self-diffusion behavior in a given state is described
by the covariance matrix in the above likelihood function: 

\begin{equation}
\boldsymbol{\varSigma}=\left[\begin{array}{cc}
\tau D_{1} & \tau r_{12}\sqrt{D_{1}D_{2}}\\
\tau r_{12}\sqrt{D_{1}D_{2}} & \tau D_{2}
\end{array}\right]
\end{equation}
If we invert this covariance matrix, expand the terms in the exponential,
and take the logarithm of the resultant expression, we find:

\begin{equation}
\begin{array}{c}
ln\left(P\left(z_{1},z_{2}\mid x_{1}=0,x_{2}=0,\tau,I\right)\right)=\\
ln\left(A_{4}\right)+ln\left(a_{1}\left(z_{1}\right)\right)+ln\left(a_{2}\left(z_{2}\right)\right)-\frac{z_{1}^{2}D_{2}+2z_{1}z_{2}r_{12}\sqrt{D_{1}D_{2}}+z_{2}^{2}D_{1}}{4\tau D_{1}D_{2}\left(1-r_{12}^{2}\right)}
\end{array}\label{eq:log=000020Bayes=000020binary}
\end{equation}

Recognizing this is not the usual sequence of operations, see Section
4.1.1, it is useful to consider what additional prior information
we might want to include before continuing our analysis. In many applications,
a particular reference frame is chosen to simplify the final species
continuity equations. Furthermore, it is desirable to have the resultant
theory satisfy the Gibbs-Duhem relation. Consider molar diffusion
with respect to the volume average velocity, where activities satisfy
the Gibbs-Duhem relation. The isothermal, isobaric Gibbs-Duhem relation
requires:

\begin{equation}
C_{1}\frac{dln\left(a_{1}\right)}{dz}+C_{2}\frac{dln\left(a_{2}\right)}{dz}=0
\end{equation}
or

\begin{equation}
\frac{dln\left(a_{2}\right)}{dz}=-\frac{\phi_{1}}{\phi_{2}}\frac{\widetilde{V}_{2}}{\widetilde{V}_{1}}\frac{dln\left(a_{1}\right)}{dz}\label{eq:binary=000020Gibbs-Duhem}
\end{equation}
where $\phi_{i}$ are species volume fractions, and $\widetilde{V}_{i}$
are species partial molar specific volumes.

For molar diffusion with respect to the volume average velocity, we
have the following relationship between species fluxes about the present
point of interest:

\begin{equation}
\widetilde{V}_{1}J_{1}+\widetilde{V}_{2}J_{2}=\widetilde{V}_{1}C_{1}\left(v_{1}-v^{V}\right)+\widetilde{V}_{2}C_{2}\left(v_{2}-v^{V}\right)=0
\end{equation}

We expect these relations to hold on average at a macroscopic scale,
but we are focused on inferences about particular molecules at a particular
location and time. Again, in the spirit of investigation, let us assume
that we can use this relation to further restrict the allowable motions
of our molecules, which are presently at the origin of our reference
frame:

\begin{equation}
\widetilde{V}_{1}J_{1}+\widetilde{V}_{2}J_{2}=\phi_{1}\frac{0-z_{1}}{2\tau}+\phi_{2}\frac{0-z_{2}}{2\tau}=0\label{eq:binary=000020flux=000020constraint}
\end{equation}
Thus:

\begin{equation}
z_{2}=-\frac{\phi_{1}}{\phi_{2}}z_{1}\label{eq:binary=000020position=000020constraint}
\end{equation}

\eqref{eq:binary=000020position=000020constraint} constrains the
allowable class of component 2 motions relative to those of component
1. Under this constraint, the most probable past position of component
2 is determined by the most probable past position of component 1.
Differentiating \eqref{eq:log=000020Bayes=000020binary} with respect
to $z_{1}$, recognizing $z_{2}=f(z_{1}),$and then using \eqref{eq:binary=000020Gibbs-Duhem}
and \eqref{eq:binary=000020position=000020constraint} with the assumption
that the activity gradients are constant in the $(z_{1},z_{2})$ neighborhood
of the point of interest over the relevant time scale gives:

\begin{equation}
\begin{array}{c}
\frac{dln\left(a_{1}\left(z_{1}\right)\right)}{dz_{1}}+\frac{dln\left(a_{2}\left(z_{2}\right)\right)}{dz_{2}}\frac{dz_{2}}{dz_{1}}-\frac{\partial}{\partial z_{1}}\left(\frac{z_{1}^{2}D_{2}+2z_{1}z_{2}r_{12}\sqrt{D_{1}D_{2}}+z_{2}^{2}D_{1}}{4\tau D_{1}D_{2}\left(1-r_{12}^{2}\right)}\right)=\\
\left(1+\frac{\phi_{1}^{2}}{\phi_{2}^{2}}\frac{\widetilde{V}_{2}}{\widetilde{V}_{1}}\right)\frac{dln\left(a_{1}\left(z_{1}\right)\right)}{dz_{1}}-z_{1}\left(\frac{\phi_{2}^{2}D_{2}+2\phi_{1}\phi_{2}r_{12}\sqrt{D_{1}D_{2}}+\phi_{1}^{2}D_{1}}{2\tau\phi_{2}^{2}D_{1}D_{2}\left(1-r_{12}^{2}\right)}\right)=0
\end{array}
\end{equation}
Thus:

\begin{equation}
J_{1}^{V}=C_{1}\left(\hat{v}_{1}-v^{V}\right)=C_{1}\frac{-z_{1}}{2\tau}=-D\frac{dC_{1}}{dx}
\end{equation}
where:

\begin{equation}
D=\frac{D_{1}D_{2}\left(1-r_{12}^{2}\right)}{\phi_{1}^{2}D_{1}+2\phi_{1}\phi_{2}r_{12}\sqrt{D_{1}D_{2}}+\phi_{2}^{2}D_{2}}\left(\phi_{2}^{2}+\phi_{1}^{2}\frac{\widetilde{V}_{2}}{\widetilde{V}_{1}}\right)\frac{dln\left(a_{1}\right)}{dln\left(C_{1}\right)}
\end{equation}
Or in the usual form:

\begin{equation}
D=\frac{D_{1}D_{2}\left(1-r_{12}^{2}\right)}{\phi_{1}^{2}D_{1}+2\phi_{1}\phi_{2}r_{12}\sqrt{D_{1}D_{2}}+\phi_{2}^{2}D_{2}}\left(\phi_{2}^{2}+\phi_{1}^{2}\frac{\widetilde{V}_{2}}{\widetilde{V}_{1}}\right)\left(\frac{x_{1}\widetilde{V}_{1}+x_{2}\widetilde{V}_{2}}{\widetilde{V}_{2}}\right)\varGamma\label{eq:joint=000020D}
\end{equation}

This is an information theoretic expression, based on the assumptions
described above, relating the Fickian mutual diffusion coefficient
for a binary system to the component self-diffusion coefficients.
Here we see that probability theory applied as logic provides a truly
``physics informed'' inference engine. To satisfy the required condition
that the mutual diffusion coefficient approach the dilute component
self-diffusion coefficients in the pure component limits, the correlation
coefficient, $r_{12}$, must also be a function of concentration that
approaches zero in the pure component limits. Otherwise, the theory
offers no fundamental form for $r_{12}$, leaving it open for further
investigation. This is not unlike the characterization of clustering
in Carman \citep{carman1967selfdiffusion}, where the concentration
dependent clustering function must be determined independently. In
keeping with the usual notion of ``ideal'' behavior occurring when
component molecular motions are independent, we can define ideal systems
as having component movements uncorrelated across the concentration
range, $r_{12}=0$. 

Given measured self- and mutual diffusion coefficients at equivalent
concentrations, or some functional representation thereof, \eqref{eq:joint=000020D}
can be rearranged into a quadratic equation for $r_{12}$. For the
systems we have examined, we find either two real solutions for $r_{12}$,
a lesser root that is always less than or equal to zero and a greater
root that can be either negative or positive, or no real solution.
Rearranging \eqref{eq:joint=000020D} into quadratic form in $r_{12}$,
we find two conditions for real roots:
\begin{equation}
D_{1}\left(\phi_{2}^{2}+\phi_{1}^{2}\frac{\widetilde{V}_{2}}{\widetilde{V}_{1}}\right)\left(\frac{x_{1}\widetilde{V}_{1}+x_{2}\widetilde{V}_{2}}{\widetilde{V}_{2}}\right)\varGamma-D_{m}\phi_{2}^{2}\geq0\label{eq:real=000020r12s}
\end{equation}

and

\begin{equation}
D_{2}\left(\phi_{2}^{2}+\phi_{1}^{2}\frac{\widetilde{V}_{2}}{\widetilde{V}_{1}}\right)\left(\frac{x_{1}\widetilde{V}_{1}+x_{2}\widetilde{V}_{2}}{\widetilde{V}_{2}}\right)\varGamma-D_{m}\phi_{1}^{2}\geq0\label{eq:real=000020r12s-1}
\end{equation}

With regard to ideal behavior, the greater root is always zero when
the mutual is constructed from the Darken form and the self-diffusion
coefficients are inversely proportional to their associated molar
volumes. Of course, one could have ideal behavior in the Bayesian
model sense, $r_{12}=0$, even when the self-diffusion coefficients
are not inversely proportional to their associated molar volumes,
and the model would generate mutual diffusion coefficients that differ
from the ``ideal'' Darken values. When there is no real solution,
the model is indicating either errant measurements; inconsistent characterization
of diffusion and/or thermodynamic behavior; or some other missing
physics, e. g., clustering, in the model. Lack of a real solution
is a failure mode that should spark further research. In the systems
examined below, we observe such failure when strong clustering is
known or expected to occur. As we will see below, the model fails
to give real roots for $r_{12}$ in several systems in which strong
clustering is not expected. These failures are not fundamental insofar
as taking a Bayesian viewpoint of the uncertainty in our representations
of both diffusion coefficients and component activities allows descriptions
with real roots for $r_{12}$ to be found.

\subsubsection{Incorporating the Flux Constraint as Prior Information}

As stated above, the flux constraint, \eqref{eq:binary=000020flux=000020constraint},
should be incorporated into the probability distribution as a constraint
on expectations, not as a hard constraint on individual molecular
motions. This can be done using the Maximum Entropy procedure \citep{caticha2007information}.
The challenge in this approach is that the resultant form of the probability
distribution for the past positions of molecules becomes unwieldy,
rendering concise expressions of the mutual diffusion coefficients,
like \eqref{eq:joint=000020D}, unavailable. We reserve this line
of inquiry for future investigation.

\subsubsection{Comparison with Some Existing Theories}

To understand what constraints are imposed by their algebraic forms,
we examine the behavior of various models for the MS diffusion coefficients,
\DJ , in dimensionless terms, for some simplified systems with constant
self-diffusion coefficients. By removing the nonlinearities associated
with the concentration dependent behavior prevalent in real systems,
we can develop a clearer understanding of the constraints imposed
by the various model forms. 

In the first case, shown in Fig. 1(a), the two components have identical
molecular weights and molar volumes but the component 2 self-diffusion
coefficient is 10\% greater than that of component 1. The Darken coefficients
vary linearly across the concentration range. The Vignes coefficients
deviate below the Darken values only slightly. The Bayesian model
values, however, display a sigmoidal shape across the concentration
range that deviate below and above the Darken values and approach
the pure component limits tangentially to the dilute component self-diffusion
values. Tangential approach of the frictional term to the self-diffusion
coefficients and sigmoidal deviations from the Darken coefficients
are both behaviors that could potentially be investigated experimentally
in an ideal system and appears to be unique to the Bayesian model.
\begin{center}
\begin{figure}[H]
\begin{centering}
\begin{tabular}{cc}
\includegraphics[scale=0.5]{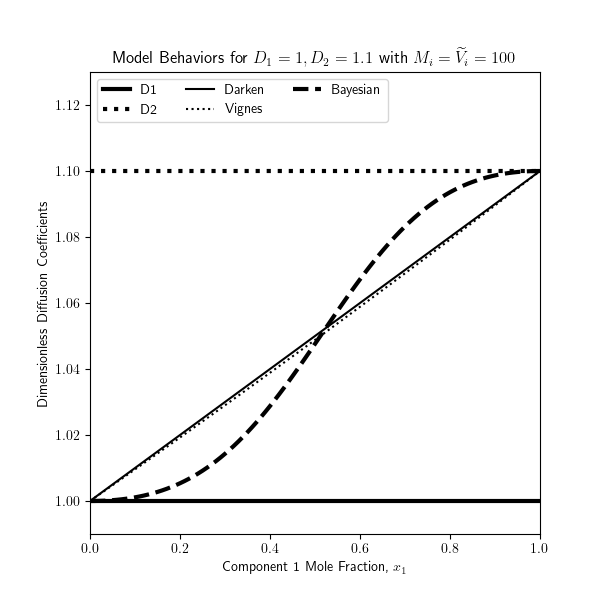} & \includegraphics[scale=0.5]{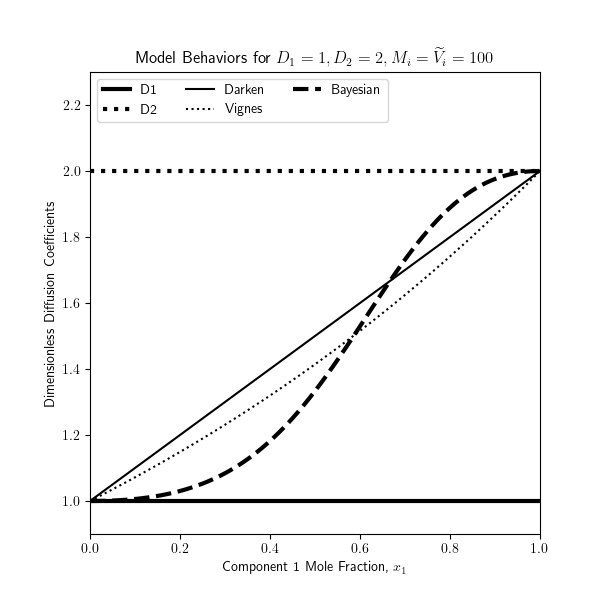}\tabularnewline
(a) $D_{2}=1.1D_{1}$ & (b) $D_{2}=2D_{1}$\tabularnewline
\end{tabular}
\par\end{centering}
\caption{Binary model behaviors for $\mkern2mu\rule[0.75ex]{0.75ex}{0.06ex}\mkern-8mu D$
with varying $D_{2}/D_{1}$. }

\end{figure}
\par\end{center}

In Figure 1(b), the self-diffusion coefficient of component 2 is increased
to double that of component 1. As the difference in self-diffusion
coefficients gets bigger, the deviations of the Vignes and Bayesian
coefficients from the Darken values grow in both magnitude and, in
the Bayesian case, asymmetry about the center of the concentration
range. The Vignes deviations are one-sided, while the Bayesian model
deviates to both sides of the Darken values. The Bayesian model deviates
below the Darken values over a fraction of the concentration range
in approximate proportion to the ratio of the greater to lesser self-diffusion
coefficients and above the Darken values over the remaining concentration
range.

In the first two cases, we varied the self-diffusion coefficients
independently of the molar volumes, but some relationship between
the two is expected. Fig. 2 shows the results when the molar volume
of the second component is set so that the ratio of component self-diffusion
coefficients is inversely proportional to the ratio of component molar
volumes. In this case, the sigmoidal behavior of the Bayesian model
disappears and the Bayesian values coincide with the Darken values. 

In fact, one can show algebraically that, when the ratio of the self-diffusion
coefficients is equal to the inverse ratio of molar volumes:

\begin{equation}
\frac{D_{2}}{D_{1}}=\frac{\widetilde{V}_{1}}{\widetilde{V}_{2}},\label{eq:self-ratio}
\end{equation}
the Bayesian and Darken models are equivalent. The ratio in \eqref{eq:self-ratio}
also arises when Bearman's \citep{bearman1961onthe} friction-based
theory is applied to regular solutions. In that case, Bearman's theory
results in an MS diffusion coefficient, \DJ , that takes the Darken
form.

\begin{figure}[H]
\centering{}\includegraphics[scale=0.5]{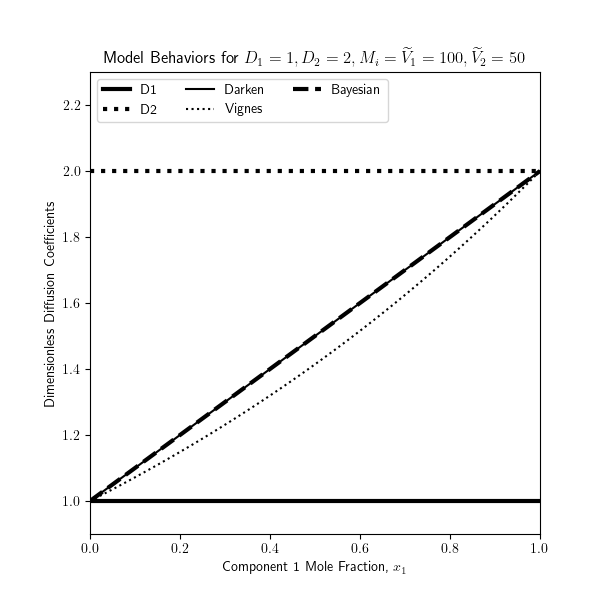}\caption{Binary model behaviors for $\mkern2mu\rule[0.75ex]{0.75ex}{0.06ex}\mkern-8mu D$
with $D_{1}/D_{2}=\widetilde{V}_{2}/\widetilde{V}_{1}$.}
\end{figure}

The results above do not consider associations between component molecules.
To gain insight into the effects of such associations, we start with
a generalized Darken form based on Carman's \citep{carman1967selfdiffusion}
model where clusters may form in either component:

\begin{equation}
\mkern2mu\rule[0.75ex]{0.75ex}{0.06ex}\mkern-8mu D=x_{1}f_{2}D_{2}+x_{2}f_{1}D_{1}\label{eq:cluster=000020Darken}
\end{equation}
The $f_{i}$ are composition dependent functions included to account
for clustering of component $i$. For example, \citet{moggridge2012prediction}
uses a fixed $f_{1}=2$ to represent dimerization of the alcohols
in solutions of methanol or ethanol with carbon tetrachloride. For
the mutual diffusion coefficients to approach their appropriate pure
component limits, must have:

\begin{equation}
\begin{array}{cc}
\underset{x_{2}\rightarrow1}{lim}f_{1}=1, & \underset{x_{1}\rightarrow1}{lim}f_{2}=1\end{array}
\end{equation}

It is clear that as a general weighted function, \eqref{eq:cluster=000020Darken}
could be used to fit any continuous functional representation of \DJ{}
given continuous functional representations of the component self-diffusion
coefficients. To retain their physical significance as adjustments
for intra-species clustering, the $f_{i}$ must take values equal
to or greater than 1. Given that constraint, it is instructive to
see how the cluster weighting functions impact \DJ , if only for comparison
with the Bayesian model results discussed below. 

To model a function that obeys the required limits and is broadly
active over the concentration range, let $f_{i}$ take the form:

\begin{equation}
\begin{array}{c}
f_{i}=1+f\\
f=f_{i}^{'}\left(4x_{1}x_{2}\right)^{0.25}
\end{array}\label{eq:cluster=000020f}
\end{equation}

Figure 3 shows the cluster model results when $f^{'}$ ranges from
0 to 1, covering the range from monomers to dimers for component 1
in Fig. 3a and for component 2 in Fig. 3b. The mole fraction weighting
of the cluster factor results in increasing enhancement of \DJ{} as
the concentration of the non-clustering component increases up to
the point where f decays as it approaches the pure component limits. 

The inclusion of an explicit cluster size coefficient is a strong
constraint on the modeled process, scaling the corresponding self-diffusion
coefficient directly. The correlation coefficient in the self-diffusion
covariance matrix of the Bayesian model is an alternative approach
to account for component interactions. Figure 4 shows the behavior
of the model system where we use the expression for $f$ in \eqref{eq:cluster=000020f}
to represent the concentration dependence of the correlation coefficient,
$r_{12}$, with $f^{'}$ spanning the range from $-0.2$ to $0.2$.
Negative correlations between species lead to positive deviations
from the Darken form, except near the pure component limits, where
the deviations become negative. Positive correlations lead to negative
deviations from the Darken form that are more broadly distributed
relative to the positive deviations at equivalent negative correlation
values.

\begin{figure}[H]

\centering{}%
\begin{tabular}{cc}
\includegraphics[scale=0.5]{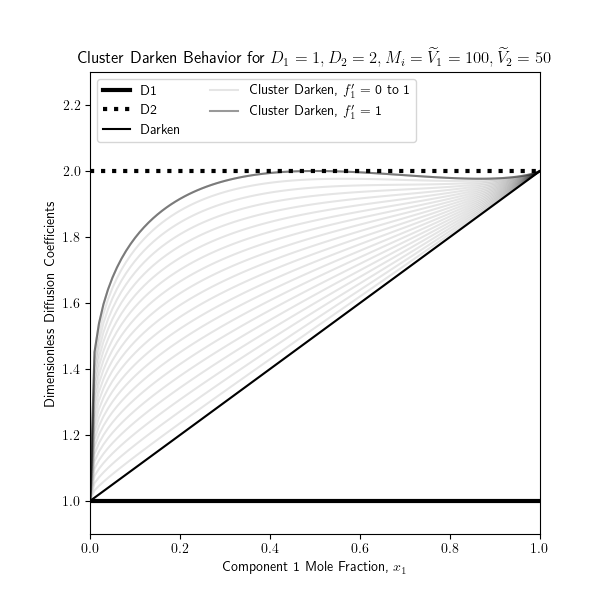} & \includegraphics[scale=0.5]{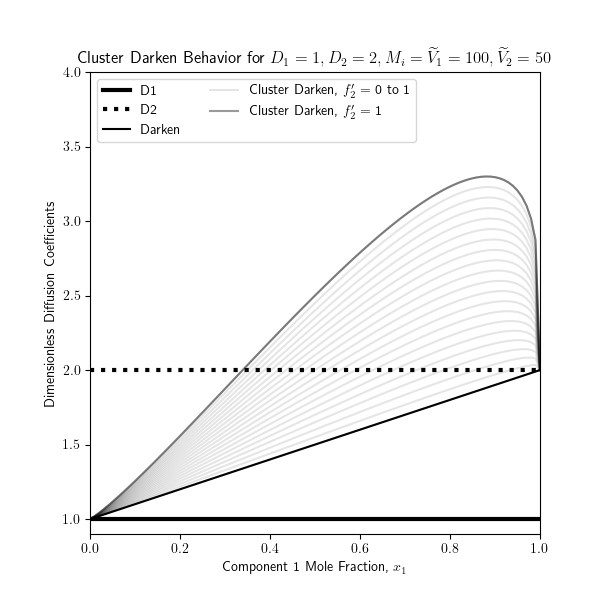}\tabularnewline
(a) $f_{1}\in(1,2),f_{2}=1$ & (b) $f_{1}=1,f_{2}\in(1,2)$\tabularnewline
\end{tabular}\caption{Behavior of cluster Darken model, \eqref{eq:cluster=000020Darken},
with varying cluster coefficients.}
\end{figure}

\begin{figure}[H]
\centering{}%
\begin{tabular}{cc}
\includegraphics[scale=0.5]{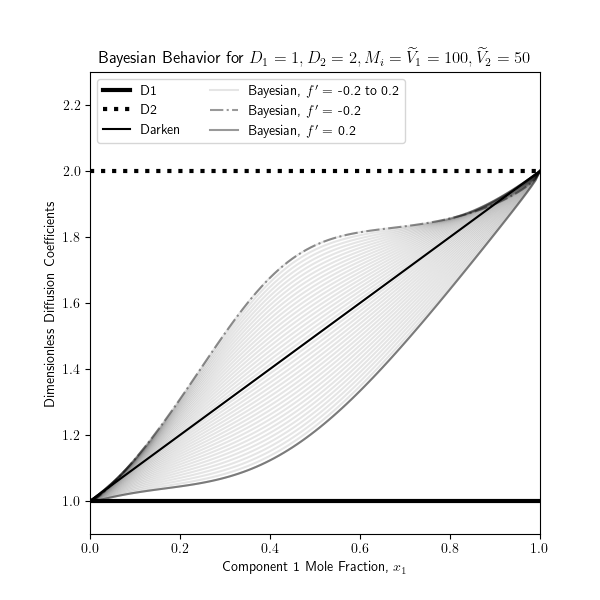} & \includegraphics[scale=0.5]{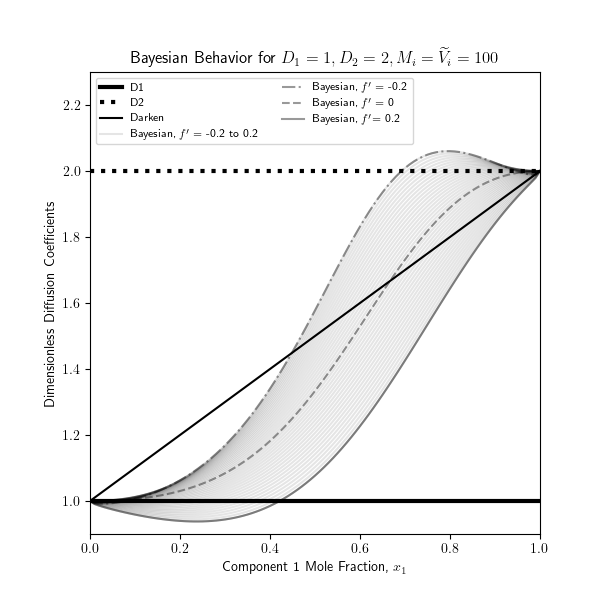}\tabularnewline
(a) $f'\in(-0.2,0.2),D_{1}/D_{2}=\widetilde{V}_{2}/\widetilde{V}_{1}$ & (b) $f'\in(-0.2,0.2),D_{1}/D_{2}\neq\widetilde{V}_{2}/\widetilde{V}_{1}$\tabularnewline
\end{tabular}\caption{Behavior of Bayesian model, \eqref{eq:joint=000020D}, with varying
correlations.}
\end{figure}

The case studies above make clear that the clustering and correlation
functions, $f_{i}$ and $r_{12}$, considered as general fitting functions,
could be used to shape the predicted \DJ{} over a broad spectrum of
behavior. The Bayesian model is distinct in its ability to account
for negative deviations from the Darken form when inter-species correlations
are positive. The clustering model, on the other hand, allows for
stronger positive deviations from the Darken form than are available
to the Bayesian model given the range of allowable correlation coefficients.
Neither the cluster function nor the correlation function is commonly
determined experimentally. Using them as fitting functions may, however,
provide insights into the observed mutual diffusion behavior of specific
systems.

\subsubsection{Comparison with Published Binary Diffusion Data}

Moggridge \citeyearpar{moggridge2012prediction2} presented an extensive
comparison between experimental data, the Darken equation, and his
thermodynamically adjusted model:

\begin{equation}
D=(x_{1}D_{2}+x_{2}D_{1})\Gamma^{\alpha}\label{eq:Moggridge=0000202}
\end{equation}
Here, we consider a subset of those binary systems where the measured
mutual diffusion data show both positive and negative deviations from
the Darken predictions, with varying degrees of expected cluster behavior.
In each case, we performed a least-squares fit of the published self-
and mutual diffusion data using polynomial finite element basis functions
\citep{dhatt1984thefinite} such that the pure component limits of
the mutual diffusion function match the appropriate self-diffusion
limits. The thermodynamic factor, $\Gamma$, for each system was computed
using Wilson's model \citep{wilson1964vaporliquid} for the activity
coefficients with parameters from the DECHEMA vapor-liquid equilibrium
data collection \citep{gmehling1993vapourliquid}. Physical parameters
for all components are given in Table 1. Wilson parameters for all
systems are given in Table 2. The Wilson parameters for diethyl ether
and chloroform are based on vapor-liquid equilibrium data measured
at 333 K but were applied in analysis after scaling to 298 K, in agreement
with the diffusion data.

\begin{table}[H]
\caption{Component physical properties \citep{haynes2016crchandbook} .}

\centering{}%
\begin{tabular}{ccc}
Component & Molecular Weight ($g/mol$) & Density ($g/cm^{3})$\tabularnewline
\hline 
\hline 
Acetone & 58.079 & 0.7845\tabularnewline
Benzene & 78.112 & 0.8765\tabularnewline
Chloroform & 119.378 & 1.4788\tabularnewline
Cyclohexane & 84.159 & 0.7739\tabularnewline
Diethyl Ether & 74.121 & 0.7138\tabularnewline
n-Hexane & 86.175 & 0.6606\tabularnewline
Water & 18.015 & 0.997\tabularnewline
\end{tabular}
\end{table}

\begin{table}[H]
\caption{Binary systems Wilson parameters \citep{gmehling1993vapourliquid}.}

\centering{}%
\begin{tabular}{cccc}
System & T (K) & $A_{12}$ & $A_{21}$\tabularnewline
\hline 
\hline 
Cyclohexane - Benzene & 298 & 133.751 & 170.4476\tabularnewline
n-Hexane - Benzene & 298 & 237.6292 & 225.595\tabularnewline
Acetone - Benzene & 298 & 418.0568 & -104.7155\tabularnewline
Acetone - Water & 298 & -35.189 & 1468.9208\tabularnewline
Acetone - Chloroform & 298 & -61.812 & -431.5877\tabularnewline
Diethyl Ether - Chloroform & 333 & -172.3024 & -285.9477\tabularnewline
Diethyl Ether - Chloroform, temperature adjusted & 298 & -154.2007 & -255.9067\tabularnewline
\end{tabular}
\end{table}

For each system, we present three figures. In the first, we compare
experimental data to mutual diffusion coefficients computed from:
the Darken model, \eqref{eq:Moggridge=0000202}, with $\alpha=1$;
Moggridge's \citep{moggridge2012prediction2} thermodynamically modified
Darken form, \eqref{eq:Moggridge=0000202} with $\alpha=0.64$; the
Bayesian model, \eqref{eq:joint=000020D}, with the correlation function,
$r_{12},$ set to zero; and the Bayesian model, \eqref{eq:joint=000020D},
with a fitted correlation function. For several systems, we also show
fitted cluster models based on \eqref{eq:cluster=000020Darken} with
an unmodified $\Gamma$. Following \citet{moggridge2012prediction2},
we show the published mutual diffusion coefficient data as points
for comparison. Published self-diffusion coefficients are also shown
as point values together with their associated finite element basis
function fit. The fitted self-diffusion polynomials are used to compute
the various model mutual diffusion coefficients across the concentration
range. The first figure is also shaded below the real root limits,
\eqref{eq:real=000020r12s} and \eqref{eq:real=000020r12s-1}, for
the mutual diffusion coefficients. Real roots can not be found when
the mutual diffusion coefficient fit is above this region. The second
figure for each system shows the fitted self-diffusion coefficients
together with various model MS diffusion coefficients, $\mkern2mu\rule[0.75ex]{0.75ex}{0.06ex}\mkern-8mu D$,
and a functional representation of $\mkern2mu\rule[0.75ex]{0.75ex}{0.06ex}\mkern-8mu D$
generated by dividing the mutual diffusion data's finite element basis
function fit by the thermodynamic factor $\Gamma$. The third figure
shows the correlation function, $r_{12}$, as regressed from the polynomial
fits to the self- and mutual diffusion coefficient data and, where
possible, as roots to \eqref{eq:joint=000020D}. For some systems,
the figure also includes the cluster function fits from the polynomial
fits to the self- and mutual diffusion coefficients. These functions
provide some insight into the deviations from ideal behavior that
occur in each system. Both the correlation and cluster function fits
are forced to zero at the pure component limits.

\paragraph*{\textit{Cyclohexane - Benzene}}

Self-diffusion coefficients for cyclohexane and benzene at 298 K are
from \citet{mills1965theintradiffusion} and digitized from Figure
1 in \citet{mccall1966selfdiffusion}. Mutual diffusion coefficients
are from \citet{rodwin1965diffusion}, \citet{sanni1971diffusion},
and \citet{sanni1973diffusivities}. For this system, the ratios of
self-diffusion coefficients are close to inverse proportion to the
ratio of species molar volumes, resulting in ideal ($r_{12}=0$ )
Bayesian MS diffusion coefficients that are close to the Darken form.
In Figure 5(b), we see that the mutual diffusion fit implied MS diffusion
coefficients exhibit positive deviations from ideal behavior, either
Darken or Bayesian, over most of the concentration range. Such deviations
can be accounted for by shrinking the thermodynamic factor \citep{moggridge2012prediction2},
with a small negative correlation coefficient, or by intra-species
clustering in either component. For this system, the required correlation
correction, shown in Figure 5(c), is a small negative correlation
over most of the concentration range. At the low end of the cyclohexane
concentration range, the real correlation constraint \eqref{eq:real=000020r12s}
is violated, and only a best fit value can be found. Failure to find
a real correlation coefficient at low concentrations appears more
likely due to the mis-characterized values of the self- and mutual
diffusion coefficients than to the need for an extreme correlation
coefficient. We consider this failure further below. \citet{tomza2019tracking}
studied clustering behavior in cyclohexane and benzene, reporting
evidence of intra-species clusters that reach their maximum concentrations
in the pure component limits and inter-species clusters across the
concentration range that reach their maximum concentration in the
middle of the concentration range, with a 60/40 ratio of intra- and
inter-species clusters in the middle of the concentration range.
\begin{figure}[H]
\begin{centering}
\begin{tabular}{ccc}
\includegraphics[scale=0.4]{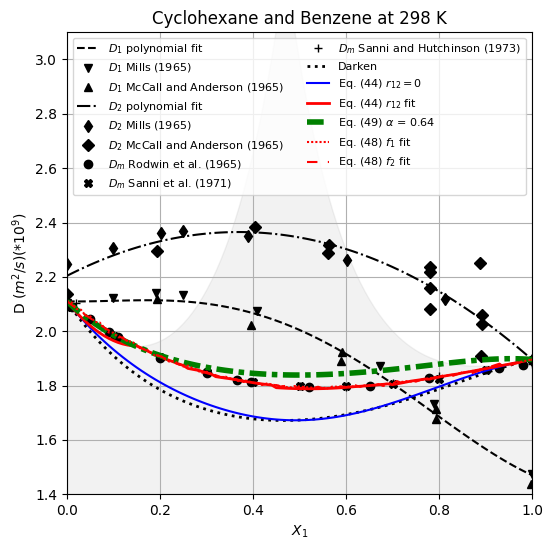} & \includegraphics[scale=0.4]{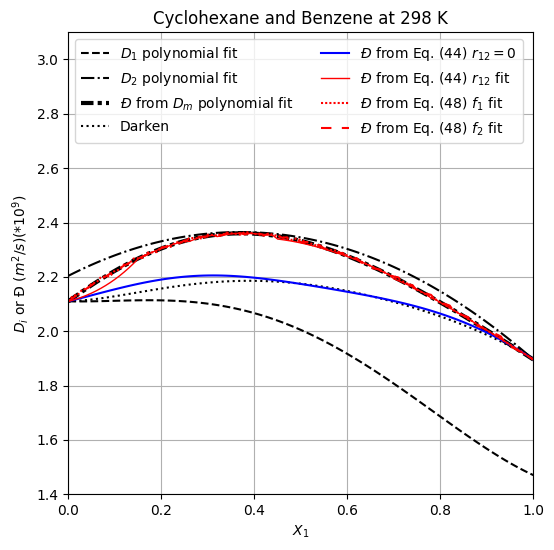} & \includegraphics[scale=0.4]{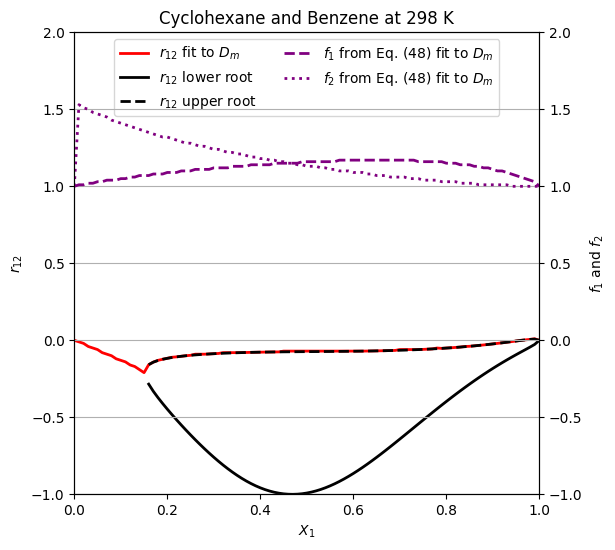}\tabularnewline
\multirow{1}{*}{(a) Self- and Mutual Diffusion Coefficients} & (b) Self- and MS Diffusion Coefficients & (c) Correlation and Clustering Coefficients\tabularnewline
\end{tabular}
\par\end{centering}
\caption{Model behavior for cyclohexane and benzene.}
\end{figure}

\paragraph*{\textit{n-Hexane - Benzene}}

Self- and mutual diffusion coefficients for n-hexane and benzene at
298 K are from \citet{harris1970mutualand}. In this case, the self-diffusion
coefficients are further from inverse proportion to the pure component
molar volumes, and the ideal Bayesian MS diffusion coefficients display
stronger sigmoidal deviation from the Darken coefficients, as described
in Section 4.1.2. The mutual diffusion fit implied MS diffusion coefficients
show significant positive deviations from the ideal behaviors. These
deviations can be accounted for with either Moggridge's adjusted the
thermodynamic factor or clustering in either component, but the system
violates the real correlation constraints in \eqref{eq:real=000020r12s}
and \eqref{eq:real=000020r12s-1} at the ends of the concentration
range. As with cyclohexane and benzene, intra- and inter-species clusters
have been observed for this system by \citet{tomza2019tracking},
with similar concentration dependence to that described above.
\begin{figure}[H]
\begin{centering}
\begin{tabular}{ccc}
\includegraphics[scale=0.4]{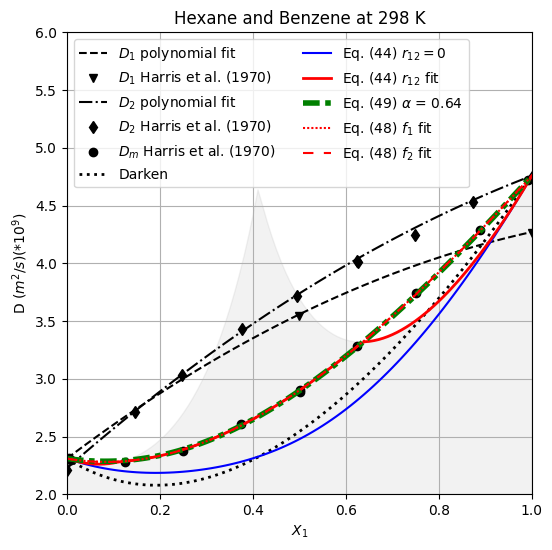} & \includegraphics[scale=0.4]{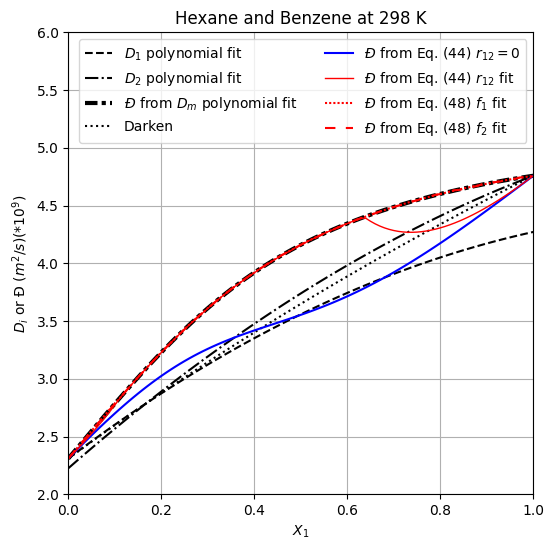} & \includegraphics[scale=0.4]{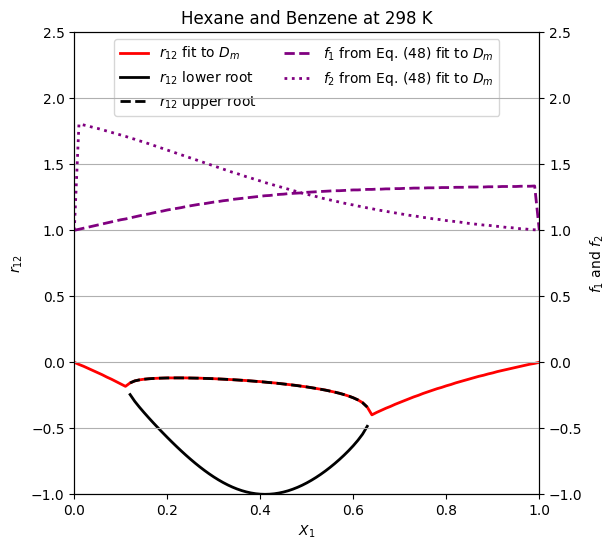}\tabularnewline
(a) Self- and Mutual Diffusion Coefficients & (b) Self- and MS Diffusion Coefficients & (c) Correlation and Clustering Coefficients\tabularnewline
\end{tabular}
\par\end{centering}
\caption{Model behavior for n-hexane and benzene.}
\end{figure}

\paragraph{\textit{Acetone - Benzene}}

Self-diffusion coefficients for acetone and benzene at 298 K are from
\citet{kamei1970selfdiffusion}. Mutual diffusion data are from \citet{anderson1958mutualdiffusion},
\citet{cullinan1965diffusion} and \citet{berg2007diffusion}. For
this system, the ideal Bayesian MS diffusion coefficients are very
close to the Darken values. The fitted mutual diffusion coefficient
data lead to MS diffusion coefficients that exhibit significant positive
deviations from the ideal behaviors. The mutual diffusion coefficients
can be fit with Moggridge's thermodynamically adjusted Darken form,
\eqref{eq:Moggridge}, with a slight negative correlation function,
or with mild clustering functions on either component. The latter
three fits are nearly indistinguishable in Figure 7(a). However, the
real correlation constraints, \eqref{eq:real=000020r12s} and \eqref{eq:real=000020r12s-1},
are again violated at the ends of the concentration range, where the
fitted values are the best approximation the model can provide to
the mutual diffusion data. These results are consistent with the molecular
simulations of acetone and benzene mixtures by \citet{povzar2015simpleand},
who report the tendency of acetone to form dimers. 
\begin{figure}[H]
\centering{}%
\begin{tabular}{ccc}
\includegraphics[scale=0.4]{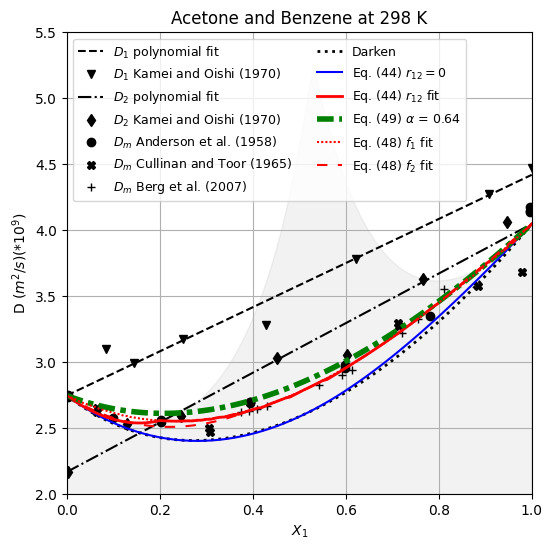} & \includegraphics[scale=0.4]{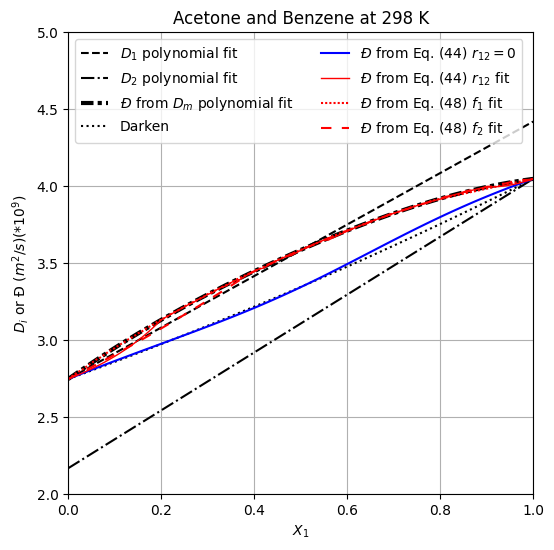} & \includegraphics[scale=0.4]{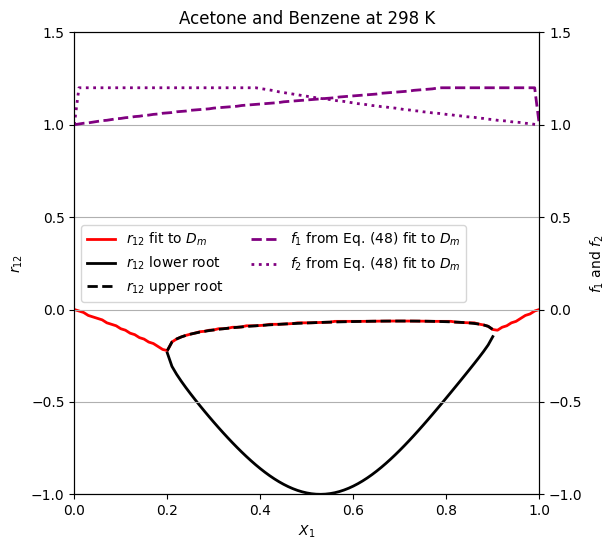}\tabularnewline
(a) Self- and Mutual Diffusion Coefficients & (b) Self- and MS Diffusion Coefficients & (b) Correlation and Clustering Coefficients\tabularnewline
\end{tabular}\caption{Model behavior for acetone and benzene.}
\end{figure}

\paragraph{\textit{Acetone - Water}}

Self-diffusion coefficients for acetone and water at 298 K are from
\citet{kamei1970selfdiffusion} and \citet{mills1980application}.
Mutual diffusion data are from \citet{anderson1958mutualdiffusion}.
For this system, the molar volume of acetone is roughly four times
that of water, yet the self-diffusion coefficients are similar over
much of the concentration range. Thus, we see significant deviations
between the ideal Bayesian MS diffusion coefficients and the Darken
values. The mutual diffusion fit implied MS diffusion coefficients
also show significant positive deviations from the ideal behaviors.
These mutual diffusion data can be reasonably represented by Moggridge's
thermodynamically adjusted model or with a fitted cluster model, but
the Bayesian model fails over all but a very small region at the low
end of the concentration range. Strong cluster functions are required
to fit the mutual diffusion data. \citet{apicella2016insights2} observed
both acetone-rich and water-rich clusters in this system.
\begin{figure}[H]
\centering{}%
\begin{tabular}{ccc}
\includegraphics[scale=0.4]{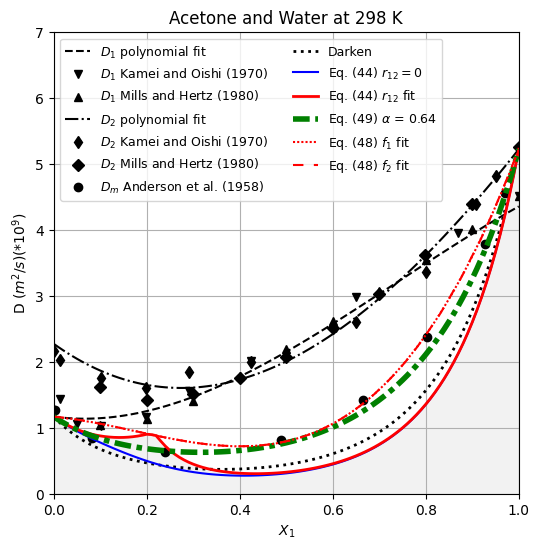} & \includegraphics[scale=0.4]{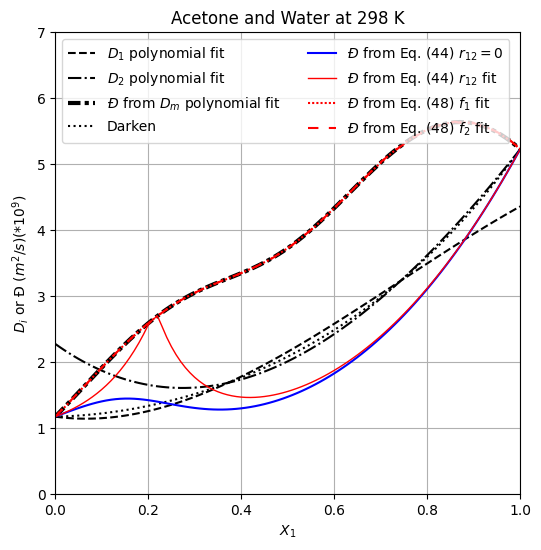} & \includegraphics[scale=0.4]{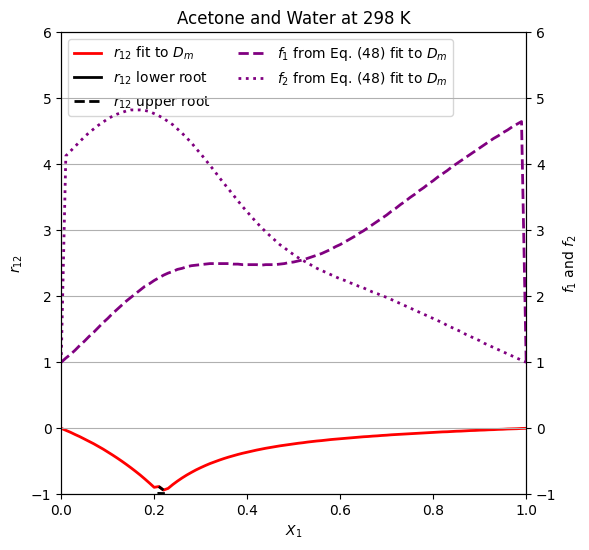}\tabularnewline
(a) Self- and Mutual Diffusion Coefficients & (b) Self- and MS Diffusion Coefficients & (b) Correlation and Clustering Coefficients\tabularnewline
\end{tabular}\caption{Model behavior for acetone and water.}
\end{figure}

\paragraph*{\textit{Acetone - Chloroform}}

Self-diffusion coefficients for acetone and chloroform at 298 K are
from \citet{dagostino2013prediction}. Mutual diffusion data are from
\citet{anderson1958mutualdiffusion} and \citet{tyn1975temperature}.
The Darken and ideal Bayesian MS diffusion coefficients for this system
are very similar. Unlike the previous systems, the MS diffusion coefficients
exhibit significant negative deviations from the ideal behaviors.
The upper Bayesian correlation coefficient accounts for the negative
deviations from ideality with positive values over the full concentration
range, excluding the pure component limits. Such behavior can not
be represented by the intra-species cluster model. This correlation
structure is consistent with the inter-species associations observed
in mixtures of acetone and chloroform reported by \citet{monakhova2014associationhydrogen}.

\begin{figure}[H]
\centering{}%
\begin{tabular}{ccc}
\includegraphics[scale=0.4]{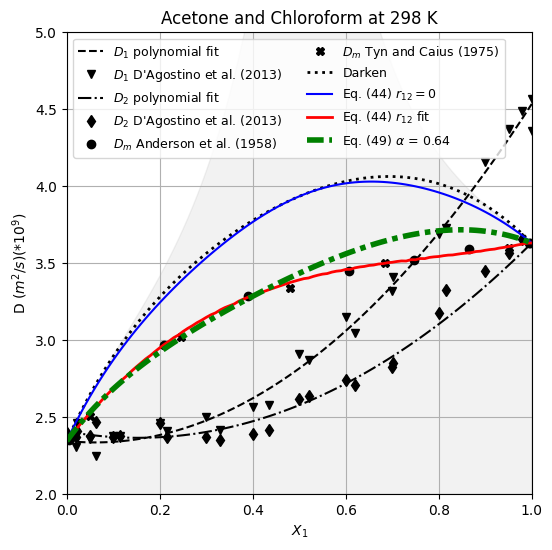} & \includegraphics[scale=0.4]{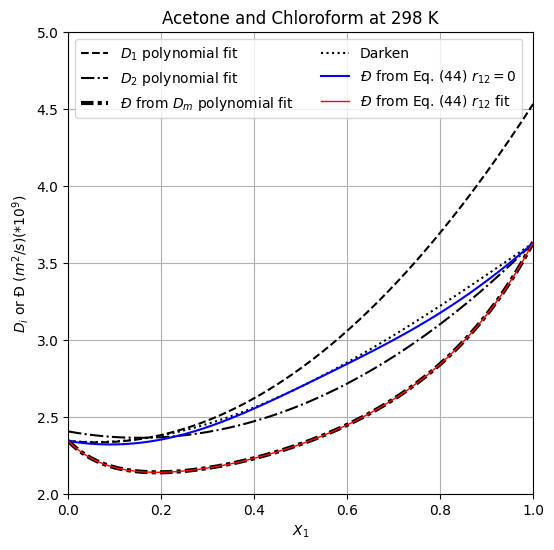} & \includegraphics[scale=0.4]{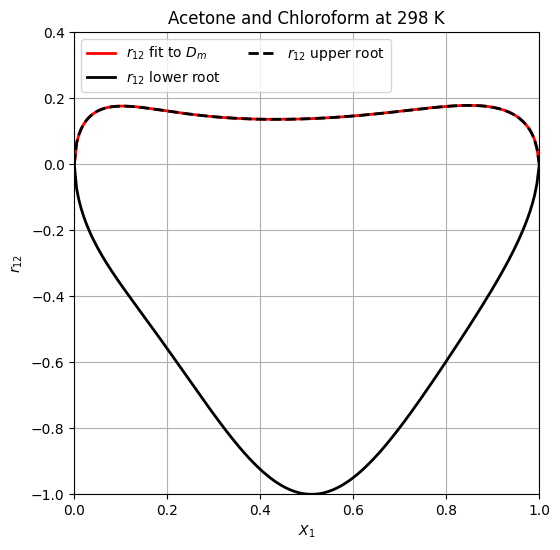}\tabularnewline
(a) Self- and Mutual Diffusion Coefficients & (b) Self- and MS Diffusion Coefficients & (b) Correlation Coefficients\tabularnewline
\end{tabular}\caption{Model behavior for acetone and chloroform.}
\end{figure}

\paragraph*{\textit{Diethyl Ether - Chloroform}}

Self- and mutual diffusion coefficient data for diethyl ether and
chloroform at 298 K were digitized from Figure 3 in \citet{weingartner1990themicroscopic}.
Additional mutual diffusion data are from \citet{anderson1961mutualdiffusion}
and \citet{sanni1973diffusivities}. The ratio of measured diethlyl
ether to chloroform self-diffusion coefficients varies significantly
over the concentration range, and so the ideal Bayesian model mutual
and MS diffusion coefficients shows the expected sigmoidal deviations
from the Darken form. The MS diffusion coefficients implied by the
polynomial fit to the mutual diffusion data and the Wilson-based $\varGamma$
show negative deviations from the Darken form at both ends of the
concentration range, and slight positive deviations in the middle.
These deviations can not be accounted for by intra-species clustering.
The Bayesian correlation function fit is positive over roughly the
lower half of the diethyl ether concentration range and becomes increasingly
negative before rising to slightly positive near the pure diethyl
ether limit. \citet{kutsyk2021mixingdynamics2} report inter-species
clusters for this system.
\begin{center}
\begin{figure}[H]
\centering{}%
\begin{tabular}{ccc}
\includegraphics[scale=0.4]{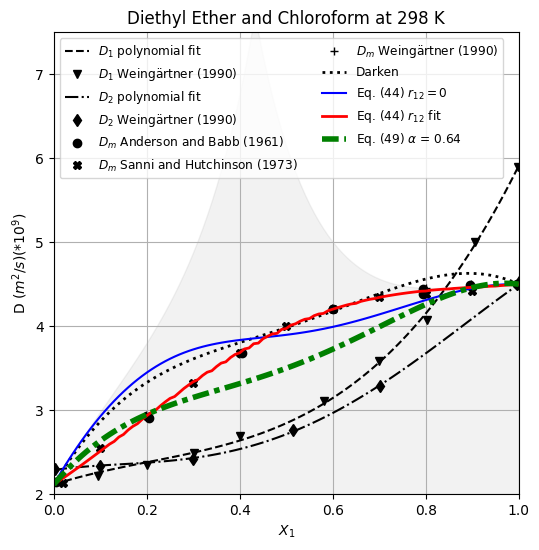} & \includegraphics[scale=0.4]{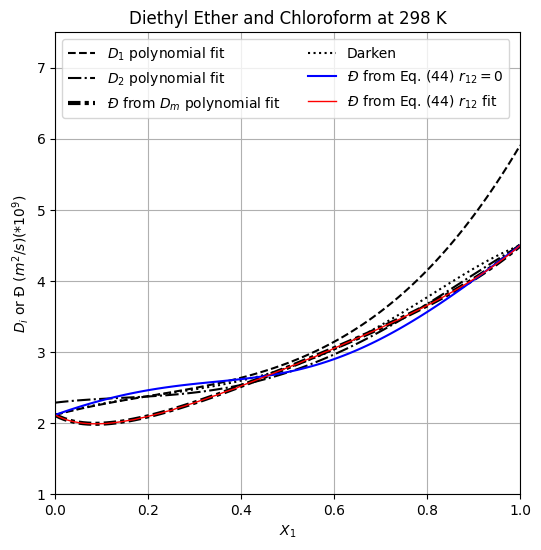} & \includegraphics[scale=0.4]{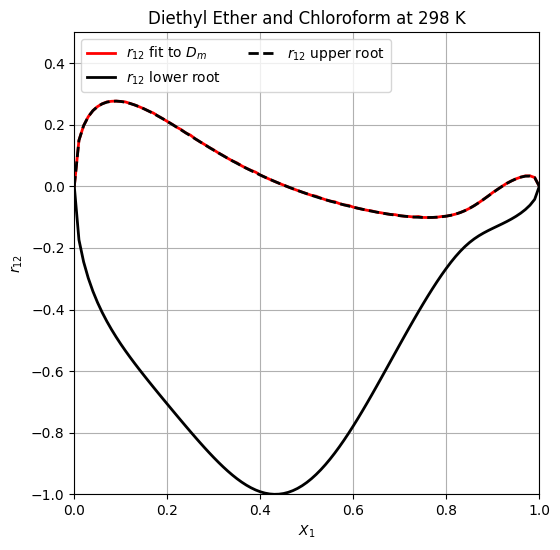}\tabularnewline
(a) Self- and Mutual Diffusion Coefficients & (b) Self- and MS Diffusion Coefficients & (b) Correlation Coefficients\tabularnewline
\end{tabular}\caption{Model behavior for diethyl ether and chloroform.}
\end{figure}
\par\end{center}

\subsubsection{Bayesian Comparisons to Published Data}

In the previous section, we took the usual approach to model comparison
with experimental data insofar as we compared results based only upon
``best fits'' to the published diffusion and vapor-liquid equilibrium
data. Following this approach, the Bayesian diffusion model yielded,
as roots, complete real correlation coefficients across the full concentration
range for only two of the six systems considered. Failure in the case
of acetone and water was not unexpected, given that system's highly
non-ideal nature. Failure in the other systems was less foreseeable.
In this section, we reconsider the systems of cyclohexane and benzene
at 298 K and acetone and benzene at 298 K, treating the underlying
data and their functional representations in a more complete Bayesian
framework. In these cases, we can sample from the posterior distributions
for the functional fits of the self- and mutual diffusion coefficients
and the activity model parameters to find model parameter sets that
yield real correlation coefficients across the full concentration
range. 

To treat the thermodynamics factor in a Bayesian way, we performed
a Bayesian regression on the data provided in the DECHEMA compilations
using the Python package Dynesty \citep{speagle2020dynesty}. The
Bayesian regression of the linked polynomial finite element basis
function representations of the self- and mutual diffusion coefficients
was also performed using Dynesty. In all cases, uniform priors were
defined on the model parameters and on the log of the standard deviations
for the Gaussian likelihood functions used in the regressions. Priors
were set to allow uncertainty in the fits beyond the standard deviation
of errors to the least-squares fits. The objective was to examine
whether the Bayesian diffusion model could provide real correlation
coefficients over the full concentration range given functional representations
of the thermodynamic and diffusion data that were reasonably consistent
with the uncertainty in the measured data. These regressions were
not meant to yield optimal Bayesian inferences on the model parameters.

For each system, we show five figures. The first figure shows a plot
of the vapor-liquid equilibrium data from the DECHEMA compilation
together with Wilson-based predictions using samples from the regression
posterior, the subset of posterior samples leading to complete real
correlations functions, and the original DECHEMA parameters. The second
figure is a scatter plot of posterior samples, samples for which the
Bayesian diffusion model gave complete real correlation functions,
and the original DECHEMA regression values. The third figure is a
plot of the self- and mutual diffusion data and polynomial representations
based on posterior samples that yield complete real correlations across
the concentration range. The fourth figure shows the least-squares
self- and MS diffusion coefficient fits and MS diffusion coefficients
computed from posterior samples that resulted in complete real correlation
functions. The final figure shows the complete real Bayesian model
correlation functions computed from the posterior samples for the
polynomial self- and mutual diffusion coefficients representations
and the Wilson parameters. 

\paragraph{\textit{Cyclohexane - Benzene}}

The x-y vapor-equilibrium data for cyclohexane and benzene, shown
in Figure 11 (a), are reasonably represented by a broad range of Wilson
parameters, shown in Figure 11 (b). The original DECHEMA Wilson parameters
are at the center of the posterior distribution, suggesting the Bayesian
regression is consistent with the original least-squares procedure.
The Wilson parameters leading to complete real correlations in the
diffusion model encompass a region of the posterior that is shifted
relative to the least-squares result. The mutual diffusion data from
multiple sources are consistent, leading to a tight distribution of
posterior mutual diffusion curves, shown in Figure 11 (c). There is
greater uncertainty in the component self-diffusion coefficients.
We find that the self-diffusion coefficients leading to complete real
correlations are evenly spread about the least-squares fits. The shifted
Wilson parameters corresponding complete real correlation functions
lead to significantly shifted MS diffusion coefficients relative to
the least-squares MS diffusion coefficients, as shown in green in
Figure 11(d). The resultant real correlation curves, shown in Figure
11 (e), are concentrated in a region between -0.06 and 0.05, with
rising values at both ends of the concentration range. As noted above,
\citet{tomza2019tracking} studied clustering behavior in cyclohexane
and benzene, reporting evidence of intra-species clusters that reach
their maximum concentrations in the pure component limits and inter-species
clusters across the concentration range that reach their maximum concentration
in the middle of the concentration range. The real correlation functions
suggest both types of clustering: positive regions at both ends correspond
to negative deviations from ideal behavior, either Darken or Bayesian,
i.e., the presence of inter-species clusters, and negative values
over the central concentration region correspond to positive deviations
from the ideal MS diffusion coefficients, indicating the presence
of intra-species clusters. The shape of the correlation functions,
however, appear reversed to Tomza et al.'s observed cluster concentration
profiles. These differences raise the question as to what extent of
any intra- or inter-species clustering is embedded in the measured
self-diffusion data. If that is the case, the Bayesian model correlation
function may be serving as a correction to the data rather than a
full, independent accounting of the system's clustering behavior.
It is also not clear how the net impact of coexisting intra- and inter-species
clustering should manifest in the correlation function. Overall, these
results indicate the sensitivity of the model's implied diffusive
correlation structure due to uncertainty in the characterizations
of thermodynamic and diffusion data. This is not surprising given
Alabi-Babalola et al.'s \citep{alabi-babalola2024rationalizing} recent
demonstration of the sensitivity of other diffusion model fits to
the choice of thermodynamic model. 

\paragraph*{\textit{Acetone - Benzene} }

For acetone and benzene, the VLE data, shown in Figure 12 (a), are
reasonably represented by the broad range of Wilson parameters, shown
in Figure 12 (b), present in the Bayesian regression posterior. The
DECHEMA Wilson parameters again appear near the center of the Bayesian
posterior, indicating consistency with the DECHEMA least-squares results.
The Wilson parameters leading to real correlation coefficients lie
directly adjacent to the DECHEMA parameters. There is greater uncertainty
in the characterizations of the mutual diffusion behavior for this
system relative to that of cyclohexane and benzene, as seen in Figure
12 (c). The acetone self-diffusion samples that lead to real correlations
are biased above the least-squares fit, while those for benzene are
more evenly distributed around the least-squares result. The MS diffusion
coefficients implied by the posterior samples, shown in Figure 12
(d), show predominantly negative deviations from ideal behavior, in
marked contrast to the least-squares implied values, which show positive
deviations from ideal behavior. The posterior sampled real correlations
are predominantly positive though small, generally less than 0.05
across the concentration range, suggesting some degree of inter-species
association. This is appears to be inconsistent with the molecular
simulations of acetone and benzene mixtures of \citet{povzar2015simpleand},
who report the tendency of acetone to form dimers but no inter-species
clusters. 

\begin{figure}[H]
\centering{}%
\begin{tabular}{ccc}
\includegraphics[scale=0.35]{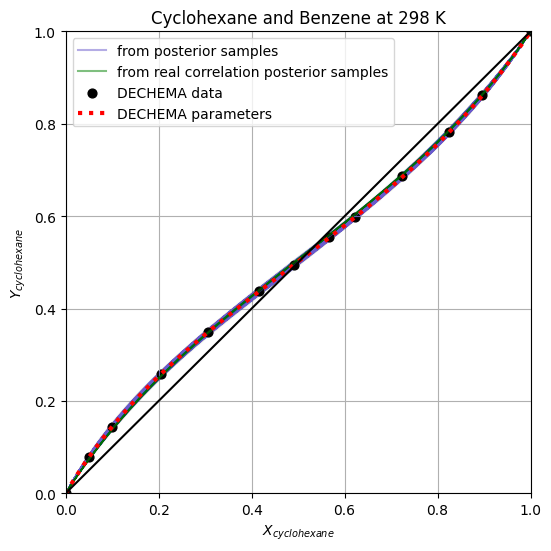} &  & \includegraphics[scale=0.35]{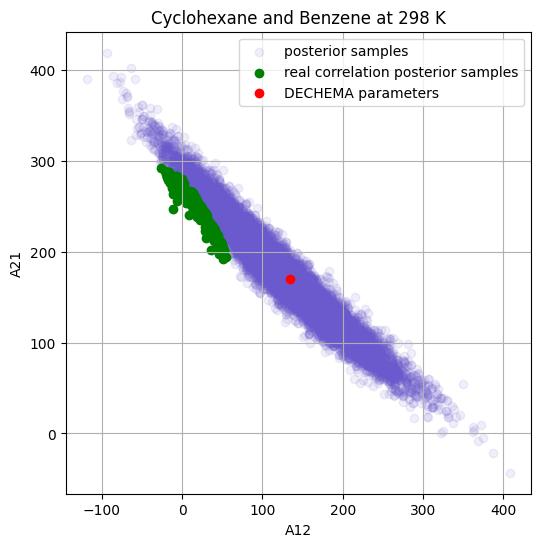}\tabularnewline
(a) VLE Data and Wilson Predictions &  & (b) Wilson Parameters\tabularnewline
\includegraphics[scale=0.35]{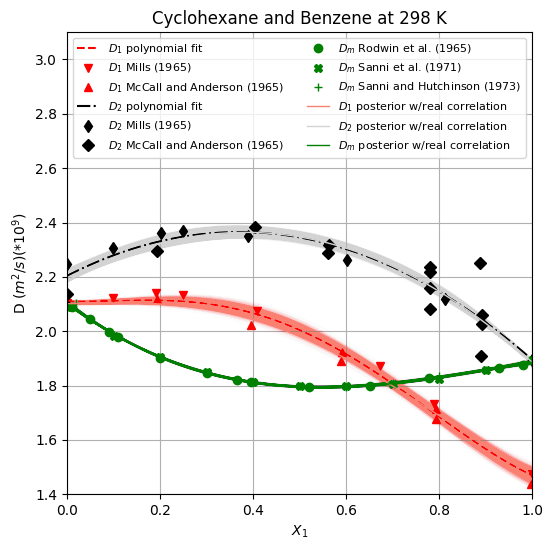} & \includegraphics[scale=0.35]{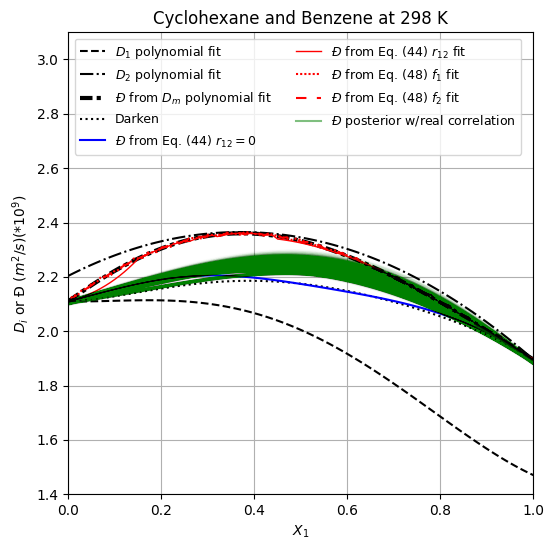} & \includegraphics[scale=0.35]{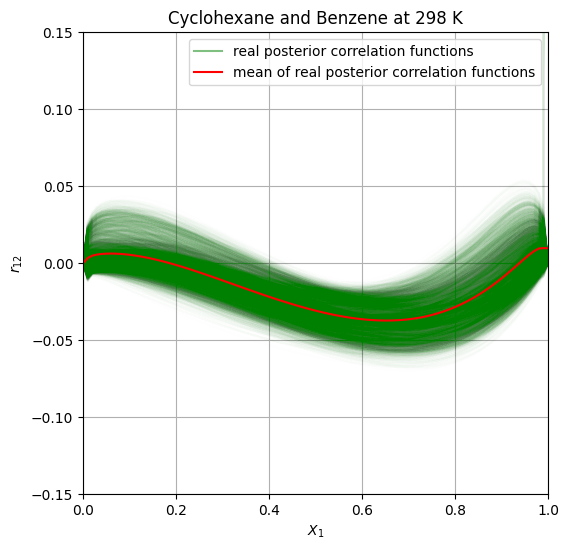}\tabularnewline
(c) Self- and Mutual Diffusion Coefficients & (d) Self- and MS Diffusion Coefficients & (e) Correlation Coefficients\tabularnewline
\end{tabular}\caption{Posterior views for cyclohexane and benzene.}
\end{figure}
\begin{figure}[H]
\centering{}%
\begin{tabular}{ccc}
\includegraphics[scale=0.35]{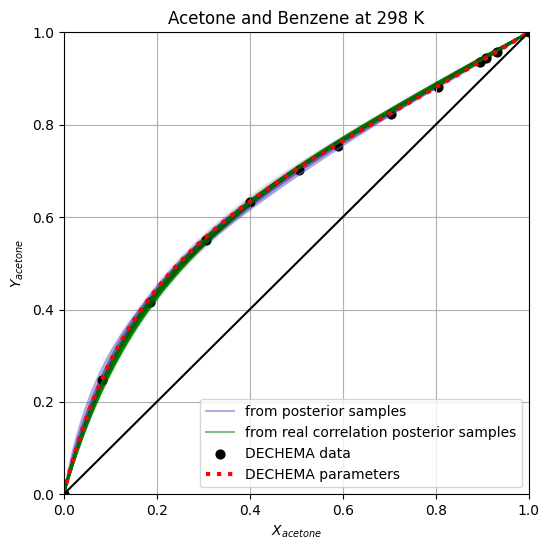} &  & \includegraphics[scale=0.35]{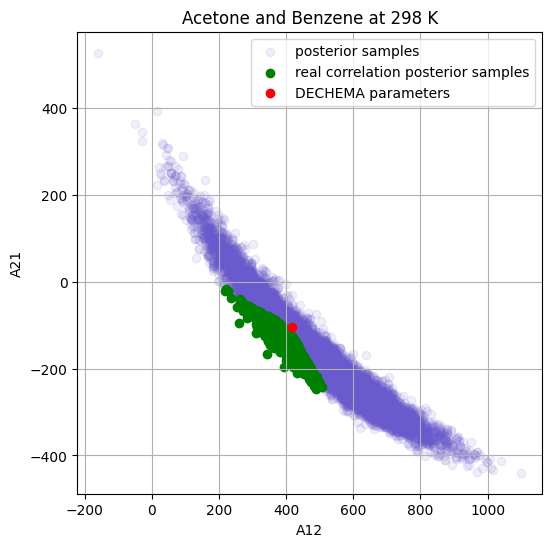}\tabularnewline
(a) VLE Data and Wilson Predictions &  & (b) Wilson Parameters\tabularnewline
\includegraphics[scale=0.35]{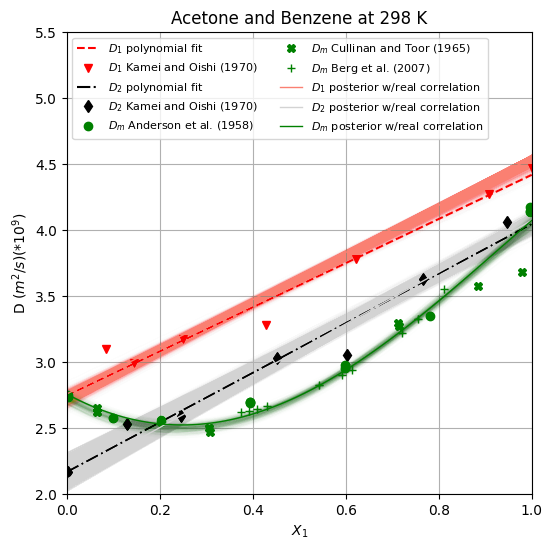} & \includegraphics[scale=0.35]{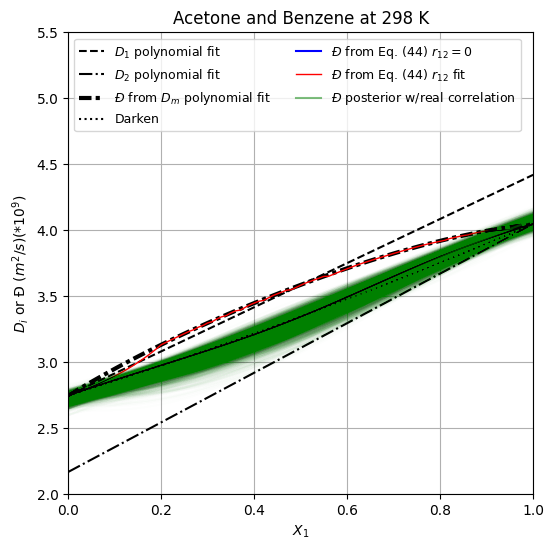} & \includegraphics[scale=0.35]{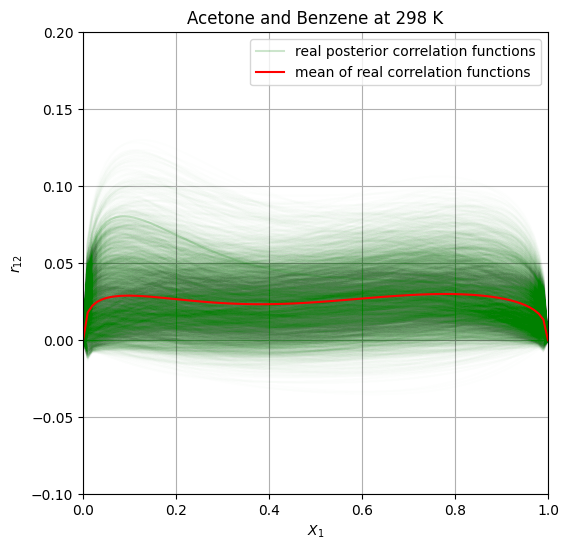}\tabularnewline
(c) Self- and Mutual Diffusion Coefficients & (d) Self- and MS Diffusion Coefficients & (e) Correlation Coefficients\tabularnewline
\end{tabular}\caption{Posterior views for acetone and benzene.}
\end{figure}

\paragraph*{\textit{Other Systems} }

Similar analysis of the systems n-hexane and benzene, and acetone
and water, failed to yield real correlation coefficients in the Bayesian
diffusion model across the full concentration range. Both of these
systems required stronger clustering functions to fit their mutual
diffusion data. While the model can provide a physically reasonable
description of four of the six systems considered, lack of complete
characterizations for these two systems suggests the need for further
prior knowledge in construction of a model that can represent the
full range of potential correlation and cluster behavior.

\subsubsection{Correlations from Local Compositions}

Local composition-based activity models have found widespread success
in thermodynamic applications \citep{kontogeorgis2010activity}. As
noted earlier, a number of authors \citep{li2001amutualdiffusioncoefficient,zhou2013localcomposition,zhu2015alocal}
have reported improvements in diffusion model predictions when local
compositions are substituted for bulk compositions. Since local compositions
differ from bulk compositions as a result of intermolecular associations,
they may provide a starting point for \textit{a priori} estimates
of the Bayesian model correlation function for some systems. If we
assume that the inter-species local mole fractions in positive excess
of the corresponding bulk value represent associations that move with
correlation 1, and those in negative excess to the bulk value represent
associations that move with correlation -1, we can posit that a weighted
blend of those excess values represents the average inter-species
correlation: 

\begin{equation}
r_{12}=x_{1}\left(x_{21}-x_{2}\right)+x_{2}\left(x_{12}-x_{1}\right)\label{eq:local=000020composition=000020r12}
\end{equation}

where $x_{21}$ represents the local mole fraction of component 2
around a molecule of component 1, and $x_{12}$ represents the local
mole fraction of component 1 around a molecule of component 2. For
the Wilson model, these are given by the expressions:

\begin{equation}
x_{21}=\frac{x_{2}\Lambda_{21}}{x_{1}+x_{2}\Lambda_{21}},\qquad x_{12}=\frac{x_{1}\Lambda_{12}}{x_{1}\Lambda_{12}+x_{2}}\label{eq:local=000020compositions}
\end{equation}

$\Lambda_{12}$ and $\Lambda_{21}$ are given by:

\begin{equation}
\Lambda_{21}=\frac{\widetilde{V}_{1}}{\widetilde{V}_{2}}\exp\left(-\frac{A_{21}}{RT}\right),\qquad\Lambda_{12}=\frac{\widetilde{V}_{2}}{\widetilde{V}_{1}}\exp\left(-\frac{A_{12}}{RT}\right)\label{eq:gammas}
\end{equation}

where $A_{12}$ and $A_{21}$ are the Wilson parameters, as listed
in Table 2 and shown in Figures 11 and 12 for the present systems
of interest.

\paragraph{Cyclohexane and Benzene}

Figure 13 (a) shows the Bayesian model predictions for the mutual
diffusion coefficients in the systems of cyclohexane and benzene at
298 K based on samples from the Bayesian regression posteriors for
this system's self-diffusion coefficients and Wilson parameters. Figure
13 (b) shows the local composition-based correlation functions, \eqref{eq:local=000020composition=000020r12},
based on the posterior samples of the Wilson parameters used to compute
the results in Figure 13 (a). The means of the sampled results are
also shown. We find that the mean of the sample predictions for the
mutual diffusion coefficients is in reasonable agreement with the
measured data, but the local-composition correlations are somewhat
different than the real correlation functions shown in Figure 11 (e).
The local composition-based correlations, almost entirely negative
over the full concentration range, are consistent with the intra-species
clustering observed by \citet{tomza2019tracking}, but fail to indicate
the inter-species clustering that they observed reaching maximum concentration
at the center of the concentration range. The results may, however,
simply reflect the net impact of both types of clusters.

\paragraph{Acetone and Benzene}

Figure 14 (a) shows the Bayesian model predictions, and corresponding
mean, for the mutual diffusion coefficients in the systems of acetone
and benzene at 298 K based on samples from the Bayesian regression
posteriors for this system's self-diffusion coefficients and Wilson
parameters. Figure 14 (b) shows the local composition-based correlation
functions, and corresponding mean, based on the posterior samples
of the Wilson parameters used to compute the results in Figure 14
(a). We find that the mean of the sample predictions for the mutual
diffusion coefficients is again in reasonable agreement with the measured
data, but for this system, the local-composition correlations show
the opposite sign over most of the concentration range relative to
the real correlation functions shown in Figure 12 (e). Regions of
negative correlation functions computed with the local composition-based
form, \eqref{eq:local=000020composition=000020r12}, suggest intra-species
association and are consistent with the acetone dimerization observed
in molecular simulations by \citet{povzar2015simpleand}.

\begin{figure}[H]
\centering{}%
\begin{tabular}{cc}
\includegraphics[scale=0.5]{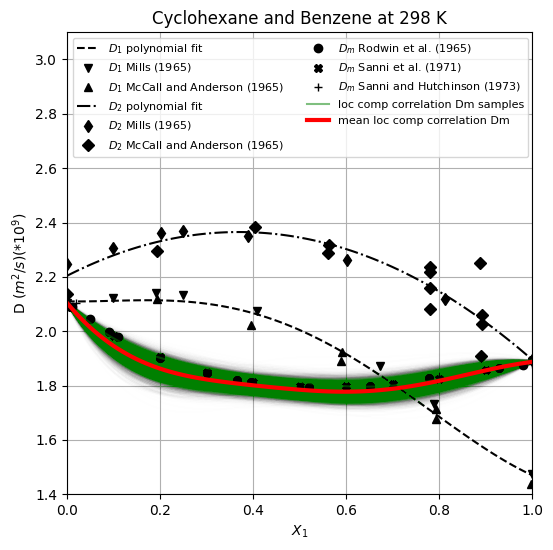} & \includegraphics[scale=0.5]{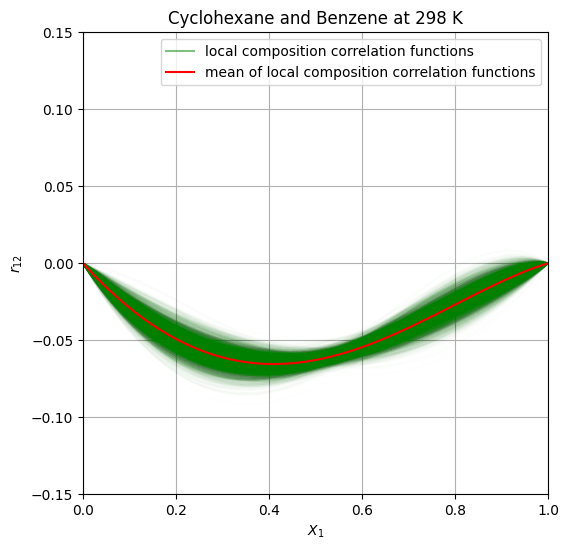}\tabularnewline
(a) Self- and Mutual Diffusion Coefficients & (b) Local Composition-based Correlation Coefficients\tabularnewline
\end{tabular}\caption{Local-composition based results for cyclohexane and benzene.}
\end{figure}

\begin{figure}[H]
\centering{}%
\begin{tabular}{cc}
\includegraphics[scale=0.5]{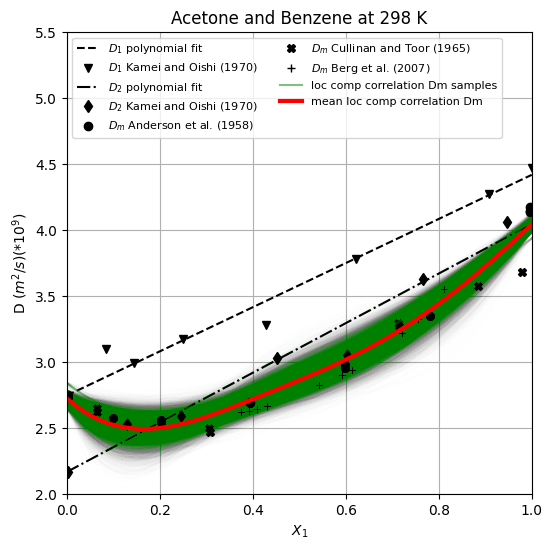} & \includegraphics[scale=0.5]{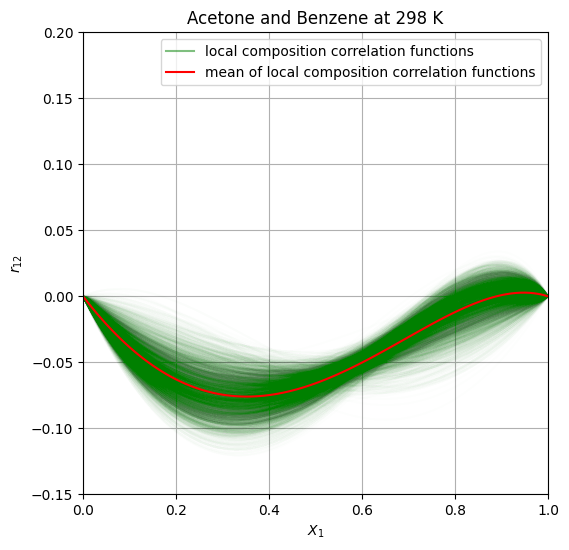}\tabularnewline
(a) Self- and Mutual Diffusion Coefficients & (b) Local Composition-based Correlation Coefficients\tabularnewline
\end{tabular}\caption{Local-composition based results for acetone and benzene.}
\end{figure}

\paragraph*{Other Systems}

A predictive form for the correlation function realizes the objective
of a theory that can predict mutual diffusion coefficients from self-diffusion
coefficients and an activity model. Applications of the thermodynamically
adjusted Darken model, \eqref{eq:Moggridge}, by Moggridge, D'Agostino
and their collaborators \citep{moggridge2012prediction,alabi-babalola2024rationalizing}
have been remarkably successful, but scaling of the thermodynamic
term in systems far from their consolute points blurs the theoretical
distinction between the inter-species frictions and the fundamental
concept that the force driving diffusive motion is the gradient in
chemical potential. Though derived on an information theoretic basis,
the Bayesian theory retains the separation of these concepts. In application,
we are hopeful that the local composition-based correlation function
will provide reasonable estimates of mutual diffusion behavior for
mildly non-ideal systems. As the intermolecular associations become
stronger and more complex, it is likely to fail, as we have observed
but not presented for the system of diethyl ether and chloroform at
298 K. More broadly, it is not clear if any of the available local
composition-based activity models can be expected to be consistent
with observed clustering behavior, even on a net intra- and inter-species
cluster basis. Predictive forms for the correlation function clearly
warrant futher development, but evaluations of such forms will have
to account for uncertainty in functional characterizations of self-diffusion
and thermodynamic parameters, which, as the results above indicate,
certainly impact perceived success of the model for any given system. 

\subsection{Multicomponent Systems}

Extension of the Bayesian diffusion model to multicomponent systems
requires only some additional algebra. We start by writing the joint
probability distribution for the past positions of the molecules of
each species presently at the position of interest, which we take
to be the origin of our chosen reference frame:

\begin{equation}
P(\boldsymbol{z}\mid\boldsymbol{x}=0,\tau,I)=A_{5}\left(\stackrel[i=1]{N}{\prod}a_{i}\left(z_{i}\right)\right)exp\left[-\frac{\boldsymbol{z}^{T}\boldsymbol{\varSigma}^{-1}\boldsymbol{z}}{2}\right]\label{eq:multicomponent=000020P}
\end{equation}
where $A_{5}$ is the normalizing constant, and:

\begin{equation}
\begin{array}{cc}
\boldsymbol{\begin{array}{c}
\boldsymbol{\sum}=\left[\begin{array}{cccc}
\tau D_{1} & \tau r_{12}\sqrt{D_{1}D_{2}} & \cdots & \tau r_{1N}\sqrt{D_{1}D_{N}}\\
\tau r_{12}\sqrt{D_{1}D_{2}} & \tau D_{2} & \cdots & \vdots\\
\vdots & \vdots & \ddots & \tau r_{N-1N}\sqrt{D_{N-1}D_{N}}\\
\tau r_{1N}\sqrt{D_{1}D_{N}} & \cdots & \tau r_{N-1N}\sqrt{D_{N-1}D_{N}} & \tau D_{N}
\end{array}\right]\\
\boldsymbol{z}=\left[\begin{array}{c}
z_{1}\\
\vdots\\
z_{N}
\end{array}\right]
\end{array}} & ,\end{array}
\end{equation}

As before, we set the derivative with respect to $\boldsymbol{z}$
of the logarithm of \eqref{eq:multicomponent=000020P} equal to zero
to find the optimal past position of the components and include additional
prior information on the component motions consistent with our chosen
reference frame and the Gibbs-Duhem equation. The former gives $z_{N}$
in terms of the other $z_{i}.$ The latter constrains the $N^{th}$
component log-activity gradients in terms of the gradients of the
other component log-activities. To solve the resultant system of equations,
we factor out the log-activity gradients and assume them locally constant
over the range spanning the individual $z_{i}$. 

A full examination of the multicomponent theory is beyond the scope
of the present paper, however, as an illustrative example, consider
the molar flux with respect to the volume average velocity. Taking
the logarithm of \eqref{eq:multicomponent=000020P} gives:

\begin{equation}
ln\left(P\left(\boldsymbol{z}\mid\boldsymbol{x}=0,\tau,I\right)\right)=ln\left(A_{5}\right)+\stackrel[i=1]{N}{\sum}\left(ln\left(a_{i}\left(z_{i}\right)\right)\right)-\frac{\boldsymbol{z}^{T}\boldsymbol{\varSigma}^{-1}\boldsymbol{z}}{2}
\end{equation}
Differentiating with respect to $z_{i}$ and imposing the flux and
Gibbs-Duhem constraints yields, for $i$ from $1$ to $N-1$:

\begin{equation}
\begin{array}{c}
\frac{dln\left(a_{i}\left(z_{i}\right)\right)}{dz_{i}}+\frac{dln\left(a_{N}\left(z_{N}\right)\right)}{dz_{N}}\frac{dz_{N}}{dz_{i}}-\boldsymbol{f}_{i}^{T}\boldsymbol{\sum}^{-1}\boldsymbol{z'}=\\
\frac{dln\left(a_{i}\right)}{dz_{i}}-\stackrel[j=1]{N-1}{\sum}\left(\frac{\phi_{i}\phi_{j}\widetilde{V}_{N}}{\phi_{N}^{2}\widetilde{V}_{j}}\frac{dln\left(a_{j}\right)}{dz_{i}}\right)-\boldsymbol{f}_{i}^{T}\boldsymbol{\sum}^{-1}\boldsymbol{z'}=0
\end{array}
\end{equation}
where:

\begin{equation}
\begin{array}{cc}
\boldsymbol{f}_{i}=\left[\begin{array}{c}
\delta_{ij}\\
\vdots\\
\delta_{ij}\\
-\frac{\phi_{i}}{\phi_{N}}
\end{array}\right] & ,\boldsymbol{z}=\left[\begin{array}{c}
z_{1}\\
\vdots\\
z_{N}
\end{array}\right]\end{array}
\end{equation}
and j indicates the row number in $\boldsymbol{f}_{i}$, i.e., $\delta_{ij}=0$
for all but the $i^{th}$ row, where $j=i$, of $\boldsymbol{f}_{i}$.
This set of equations can be cast in matrix form and solved for the
vector of optimal past positions, $\hat{\boldsymbol{z}}$:

\begin{equation}
\hat{\boldsymbol{z}}=\boldsymbol{\varLambda}^{-1}\boldsymbol{\Phi\alpha}
\end{equation}
where:

\begin{equation}
\begin{array}{ccc}
\hat{\boldsymbol{z}}=\left[\begin{array}{c}
\hat{z}_{1}\\
\vdots\\
\hat{z}_{N-1}
\end{array}\right], & \boldsymbol{\varLambda}=\boldsymbol{F}^{T}\boldsymbol{\sum}^{-1}\boldsymbol{F}, & \boldsymbol{F}=\left[\begin{array}{cccc}
1 & 0 & \cdots & 0\\
0 & 1 & 0 & \vdots\\
\vdots & 0 & \ddots & 0\\
0 & \cdots & 0 & 1\\
-\frac{\phi_{1}}{\phi_{N}} & -\frac{\phi_{2}}{\phi_{N}} & \cdots & -\frac{\phi_{N-1}}{\phi_{N}}
\end{array}\right]\end{array}
\end{equation}
and:

\begin{equation}
\begin{array}{cc}
\Phi=\left[\begin{array}{cccc}
1+\frac{\phi_{1}^{2}\widetilde{V}_{N}}{\phi_{N}^{2}\widetilde{V}_{1}} & \frac{\phi_{1}\phi_{2}\widetilde{V}_{N}}{\phi_{N}^{2}\widetilde{V}_{2}} & \cdots & \frac{\phi_{1}\phi_{N-1}\widetilde{V}_{N}}{\phi_{N}^{2}\widetilde{V}_{N-1}}\\
\frac{\phi_{2}\phi_{1}\widetilde{V}_{N}}{\phi_{N}^{2}\widetilde{V}_{1}} & 1+\frac{\phi_{2}^{2}\widetilde{V}_{N}}{\phi_{N}^{2}\widetilde{V}_{2}} & \frac{\phi_{2}\phi_{3}\widetilde{V}_{N}}{\phi_{N}^{2}\widetilde{V}_{3}} & \vdots\\
\vdots & \frac{\phi_{3}\phi_{2}\widetilde{V}_{N}}{\phi_{N}^{2}\widetilde{V}_{2}} & \ddots & \frac{\phi_{N-2}\phi_{N-1}\widetilde{V}_{N}}{\phi_{N}^{2}\widetilde{V}_{N-1}}\\
\frac{\phi_{N-1}\phi_{1}\widetilde{V}_{N}}{\phi_{N}^{2}\widetilde{V}_{1}} & \cdots & \frac{\phi_{N-1}\phi_{N-2}\widetilde{V}_{N}}{\phi_{N}^{2}\widetilde{V}_{N-2}} & 1+\frac{\phi_{N-1}^{2}\widetilde{V}_{N}}{\phi_{N}^{2}\widetilde{V}_{N-1}}
\end{array}\right], & \alpha=\left[\begin{array}{c}
\frac{dln\left(a_{1}\right)}{dz}\\
\vdots\\
\frac{dln\left(a_{N-1}\right)}{dz}
\end{array}\right]\end{array}
\end{equation}

\subsubsection{Ideal Ternary Systems}

For the case of a ternary system where the self-diffusive correlations
are all zero, solving for $\hat{z}_{1}$ and $\hat{z}_{2}$ gives:

\begin{equation}
\left[\begin{array}{c}
\hat{z}_{1}\\
\hat{z}_{2}
\end{array}\right]=\frac{2\tau}{\stackrel[i-1]{N}{\sum}\phi_{i}^{2}D_{i}}\left[\begin{array}{cc}
D_{1}\left(\phi_{2}^{2}D_{2}+\left(\phi_{1}^{2}\frac{\widetilde{V}_{3}}{\widetilde{V}_{1}}+\phi_{3}^{2}\right)D_{3}\right) & \phi_{1}\phi_{2}D_{1}\left(\frac{\widetilde{V}_{3}}{\widetilde{V}_{2}}D_{3}-D_{2}\right)\\
\phi_{2}\phi_{1}D_{2}\left(\frac{\widetilde{V}_{3}}{\widetilde{V}_{1}}D_{3}-D_{1}\right) & D_{2}\left(\phi_{1}^{2}D_{1}+\left(\phi_{2}^{2}\frac{\widetilde{V}_{3}}{\widetilde{V}_{2}}+\phi_{3}^{2}\right)D_{3}\right)
\end{array}\right]\left[\begin{array}{c}
\frac{dln\left(a_{1}\right)}{dz}\\
\frac{dln\left(a_{2}\right)}{dz}
\end{array}\right]
\end{equation}

The component flux expressions are:

\begin{equation}
J_{i}^{V}=C_{i}(\hat{v}_{i}-v^{V})=C_{i}\frac{-z_{i}}{2\tau}
\end{equation}
yielding:

\begin{equation}
\left[\begin{array}{c}
J_{1}^{V}\\
J_{2}^{V}
\end{array}\right]=\frac{-1}{\stackrel[i-1]{N}{\sum}\phi_{i}^{2}D_{i}}\left[\begin{array}{cc}
C_{1}D_{1}\left(\phi_{2}^{2}D_{2}+\left(\phi_{1}^{2}\frac{\widetilde{V}_{3}}{\widetilde{V}_{1}}+\phi_{3}^{2}\right)D_{3}\right) & C_{1}\phi_{1}\phi_{2}D_{1}\left(\frac{\widetilde{V}_{3}}{\widetilde{V}_{2}}D_{3}-D_{2}\right)\\
C_{2}\phi_{2}\phi_{1}D_{2}\left(\frac{\widetilde{V}_{3}}{\widetilde{V}_{1}}D_{3}-D_{1}\right) & C_{2}D_{2}\left(\phi_{1}^{2}D_{1}+\left(\phi_{2}^{2}\frac{\widetilde{V}_{3}}{\widetilde{V}_{2}}+\phi_{3}^{2}\right)D_{3}\right)
\end{array}\right]\left[\begin{array}{c}
\frac{dln\left(a_{1}\right)}{dz}\\
\frac{dln\left(a_{2}\right)}{dz}
\end{array}\right]
\end{equation}

To cast the result in the usual multicomponent Fickian form:

\begin{equation}
J_{i}^{V}=-\stackrel[j=1]{N-1}{\sum}D_{ij}\frac{\partial C_{j}}{\partial z},
\end{equation}
we expand the activity gradients in the component concentrations.
For a ternary system, we have:

\begin{equation}
\left[\begin{array}{c}
\frac{dln\left(a_{1}\right)}{dz}\\
\frac{dln\left(a_{2}\right)}{dz}
\end{array}\right]=\left[\begin{array}{cc}
\frac{dln\left(a_{1}\right)}{dC_{1}} & \frac{dln\left(a_{1}\right)}{dC_{2}}\\
\frac{dln\left(a_{2}\right)}{dC_{1}} & \frac{dln\left(a_{2}\right)}{dC_{2}}
\end{array}\right]\left[\begin{array}{c}
\frac{dC_{1}}{dz}\\
\frac{dC_{2}}{dz}
\end{array}\right]
\end{equation}
Grouping terms results in the following main and cross-term mutual
diffusion coefficients:

\begin{equation}
\begin{array}{c}
D_{11}=\frac{D_{1}}{\stackrel[i-1]{3}{\sum}\phi_{i}^{2}D_{i}}\left\{ \left(\phi_{2}^{2}D_{2}+\left(\phi_{1}^{2}\frac{\widetilde{V}_{3}}{\widetilde{V}_{1}}+\phi_{3}^{2}\right)D_{3}\right)\frac{dln\left(a_{1}\right)}{dln\left(C_{1}\right)}+\phi_{1}\phi_{2}\left(\frac{\widetilde{V}_{3}}{\widetilde{V}_{2}}D_{3}-D_{2}\right)\frac{dln\left(a_{2}\right)}{dln\left(C_{1}\right)}\right\} \\
D_{12}=\frac{D_{1}}{\stackrel[i-1]{3}{\sum}\phi_{i}^{2}D_{i}}\left\{ \left(\phi_{2}^{2}D_{2}+\left(\phi_{1}^{2}\frac{\widetilde{V}_{3}}{\widetilde{V}_{1}}+\phi_{3}^{2}\right)D_{3}\right)C_{1}\frac{dln\left(a_{1}\right)}{dC_{2}}+\phi_{1}\phi_{2}\left(\frac{\widetilde{V}_{3}}{\widetilde{V}_{2}}D_{3}-D_{2}\right)C_{1}\frac{dln\left(a_{2}\right)}{dC_{2}}\right\} \\
D_{21}=\frac{D_{2}}{\stackrel[i-1]{3}{\sum}\phi_{i}^{2}D_{i}}\left\{ \phi_{2}\phi_{1}\left(\frac{\widetilde{V}_{3}}{\widetilde{V}_{1}}D_{3}-D_{1}\right)C_{2}\frac{dln\left(a_{1}\right)}{dC_{1}}+\left(\phi_{1}^{2}D_{1}+\left(\phi_{2}^{2}\frac{\widetilde{V}_{3}}{\widetilde{V}_{2}}+\phi_{3}^{2}\right)D_{3}\right)C_{2}\frac{dln\left(a_{2}\right)}{dC_{1}}\right\} \\
D_{22}=\frac{D_{2}}{\stackrel[i-1]{3}{\sum}\phi_{i}^{2}D_{i}}\left\{ \phi_{2}\phi_{1}\left(\frac{\widetilde{V}_{3}}{\widetilde{V}_{1}}D_{3}-D_{1}\right)\frac{dln\left(a_{1}\right)}{dln\left(C_{2}\right)}+\left(\phi_{1}^{2}D_{1}+\left(\phi_{2}^{2}\frac{\widetilde{V}_{3}}{\widetilde{V}_{2}}+\phi_{3}^{2}\right)D_{3}\right)\frac{dln\left(a_{2}\right)}{dln\left(C_{2}\right)}\right\} 
\end{array}
\end{equation}
With the help of the Gibbs-Duhem equation, we find that when the correlation
coefficients are zero and the self-diffusion coefficients scale inversely
with the component molar volumes, the Bayesian model mutual diffusion
coefficients match those of the multicomponent Darken form, expressed
here as derived from Bearman's friction-based approach in Price and
Romdhane \citep{price2003multicomponent}:

\begin{equation}
D_{ik}=C_{i}\stackrel[j=1,j\neq i]{N}{\sum}\phi_{j}\left(D_{i}\frac{dln\left(a_{i}\right)}{dC_{k}}-D_{j}\frac{dln\left(a_{j}\right)}{dC_{k}}\right)
\end{equation}
Both main and cross diffusion coefficients also go to their correct
dilute solution limits. Further investigation of the Bayesian model
in the multicomponent context will be the subject of future research.

\section{Conclusion}

This paper presents several extensions of Jaynes' \citep{jaynes1989clearing}
derivation of Fick's Law from Bayes' Theorem. The simplest extensions
include multiple components, alternative reference frames, and a change
in the prior probability term to reflect a driving force based on
chemical potentials rather than concentrations. These extensions are
based on information about individual components in the system, yielding
model forms that have been widely used in applications to solvent-polymer
systems. The next extension is based on joint inference of molecular
motions of both components in a binary system. The unique element
of this extension is its inclusion of a composition dependent correlation
coefficient that can account for both positive and negative deviations
of the MS diffusion coefficients from ideal, i.e., independent, behavior,
owing to negative and positive, respectively, correlation between
the species self-diffusive motions. In the limit where the ratio of
component self-diffusion coefficients are in inversely proportional
to their molar volumes, the model MS diffusion coefficients equal
the Darken values. When that ratio does not hold, but inter-species
correlations are zero, the model provides a broader definition of
ideal behavior. Comparison of model fits to experimental data indicate
that the Bayesian model is capable of providing accurate correlations
of mutual diffusion behavior. While the theory provides no prescription
for computing the correlation function based on system characteristics,
we provide a possible starting point based on local compositions and
hope that having this new model relating self- and mutual-diffusion
coefficients might spur further efforts to characterize the correlation
behavior, both theoretical and experimental. We do find that the correlation
function can not account for observed behavior when strong clustering
occurs. In that sense, the model delineates systems where a more complete
description of clustering across the concentration range is required.
The final extension presents the model in its full multicomponent
form and the special case of an ideal, i.e., independent species self-diffusive
motions, ternary system. Again, the Bayesian expressions for the ternary
mutual diffusion coefficients collapse to a multicomponent form of
the Darken equation when the ratios of self-diffusion coefficients
are inversely proportional to the component molar volumes, indicating
that the Bayesian model is a generalization of the Darken model.

\section*{Acknowledgment}

An earlier, less complete, and erroneous version of the present study
was developed while the author was employed in the Corporate Research
Process Laboratory at 3M, with 3M having granted permission to publish
and present on those results. The author would like to acknowledge
the encouragement of the late Prof. L. E. Scriven to pursue development
of that then incomplete theory. His words of support were a welcomed
boost at a time when such were nowhere else forthcoming.

\section*{\textemdash\textemdash\textemdash\textemdash\textemdash\textendash{}}

\bibliographystyle{elsarticle-harv}
\addcontentsline{toc}{section}{\refname}\bibliography{InfoFick}

\begin{thebibliography}{56}
\expandafter\ifx\csname natexlab\endcsname\relax\def\natexlab#1{#1}\fi
\providecommand{\url}[1]{\texttt{#1}}
\providecommand{\href}[2]{#2}
\providecommand{\path}[1]{#1}
\providecommand{\DOIprefix}{doi:}
\providecommand{\ArXivprefix}{arXiv:}
\providecommand{\URLprefix}{URL: }
\providecommand{\Pubmedprefix}{pmid:}
\providecommand{\doi}[1]{\href{http://dx.doi.org/#1}{\path{#1}}}
\providecommand{\Pubmed}[1]{\href{pmid:#1}{\path{#1}}}
\providecommand{\bibinfo}[2]{#2}
\ifx\xfnm\relax \def\xfnm[#1]{\unskip,\space#1}\fi
\bibitem[{Alabi-Babalola et~al.(2024)Alabi-Babalola, Zhong, Moggridge and
  D'Agostino}]{alabi-babalola2024rationalizing}
\bibinfo{author}{Alabi-Babalola, O.}, \bibinfo{author}{Zhong, J.},
  \bibinfo{author}{Moggridge, G.D.}, \bibinfo{author}{D'Agostino, C.},
  \bibinfo{year}{2024}.
\newblock \bibinfo{title}{Rationalizing the use of mutual diffusion prediction
  models in non-ideal binary mixtures}.
\newblock \bibinfo{journal}{Chemical Engineering Science}
  \bibinfo{volume}{291}, \bibinfo{pages}{119930}.
\newblock \URLprefix
  \url{https://www.sciencedirect.com/science/article/pii/S0009250924002306},
  \DOIprefix\doi{https://doi.org/10.1016/j.ces.2024.119930}.
\bibitem[{Alsoy and Duda(1999)}]{alsoy1999modeling}
\bibinfo{author}{Alsoy, S.}, \bibinfo{author}{Duda, J.L.},
  \bibinfo{year}{1999}.
\newblock \bibinfo{title}{Modeling of multicomponent drying of polymer films}.
\newblock \bibinfo{journal}{AIChE Journal} \bibinfo{volume}{45},
  \bibinfo{pages}{896--905}.
\newblock \URLprefix \url{http://doi.wiley.com/10.1002/aic.690450420},
  \DOIprefix\doi{10.1002/aic.690450420}.
\bibitem[{Anderson and Babb(1961)}]{anderson1961mutualdiffusion}
\bibinfo{author}{Anderson, D.K.}, \bibinfo{author}{Babb, A.L.},
  \bibinfo{year}{1961}.
\newblock \bibinfo{title}{{Mutual} {Diffusion} {in} {Non}-{Ideal} {Liquid}
  {Mixtures}. {II}. {Diethyl} {Ether}{\textemdash}{Chloroform}}.
\newblock \bibinfo{journal}{The Journal of Physical Chemistry}
  \bibinfo{volume}{65}, \bibinfo{pages}{1281--1283}.
\newblock \URLprefix \url{https://doi.org/10.1021/j100826a001},
  \DOIprefix\doi{10.1021/j100826a001}. \bibinfo{note}{\_eprint:
  https://doi.org/10.1021/j100826a001}.
\bibitem[{Anderson et~al.(1958)Anderson, Hall and
  Babb}]{anderson1958mutualdiffusion}
\bibinfo{author}{Anderson, D.K.}, \bibinfo{author}{Hall, J.R.},
  \bibinfo{author}{Babb, A.L.}, \bibinfo{year}{1958}.
\newblock \bibinfo{title}{Mutual {Diffusion} in {Non}-ideal {Binary} {Liquid}
  {Mixtures}}.
\newblock \bibinfo{journal}{The Journal of Physical Chemistry}
  \bibinfo{volume}{62}, \bibinfo{pages}{404--408}.
\newblock \URLprefix \url{https://pubs.acs.org/doi/abs/10.1021/j150562a006},
  \DOIprefix\doi{10.1021/j150562a006}.
\bibitem[{Apicella et~al.(2016)Apicella, Li, Passaro and
  Russo}]{apicella2016insights2}
\bibinfo{author}{Apicella, B.}, \bibinfo{author}{Li, X.},
  \bibinfo{author}{Passaro, M.}, \bibinfo{author}{Russo, C.},
  \bibinfo{year}{2016}.
\newblock \bibinfo{title}{Insights on {Clusters} {Formation} {Mechanism} by
  {Time} of {Flight} {Mass} {Spectrometry}. 2. {The} {Case} of
  {Acetone}{\textendash}{Water} {Clusters}}.
\newblock \bibinfo{journal}{Journal of the American Society for Mass
  Spectrometry} \bibinfo{volume}{27}, \bibinfo{pages}{1835--1845}.
\newblock \URLprefix \url{https://pubs.acs.org/doi/10.1007/s13361-016-1464-3},
  \DOIprefix\doi{10.1007/s13361-016-1464-3}.
\bibitem[{Bearman(1961)}]{bearman1961onthe}
\bibinfo{author}{Bearman, R.J.}, \bibinfo{year}{1961}.
\newblock \bibinfo{title}{{On} {The} {Molecular} {Basis} {of} {Some} {Current}
  {Theories} {of} {Diffusion} $^{\textrm{1}}$}.
\newblock \bibinfo{journal}{The Journal of Physical Chemistry}
  \bibinfo{volume}{65}, \bibinfo{pages}{1961--1968}.
\newblock \URLprefix \url{https://pubs.acs.org/doi/abs/10.1021/j100828a012},
  \DOIprefix\doi{10.1021/j100828a012}.
\bibitem[{Berg et~al.(2007)Berg, Hansen, Shapiro and
  Stenby}]{berg2007diffusion}
\bibinfo{author}{Berg, R.W.}, \bibinfo{author}{Hansen, S.B.},
  \bibinfo{author}{Shapiro, A.A.}, \bibinfo{author}{Stenby, E.H.},
  \bibinfo{year}{2007}.
\newblock \bibinfo{title}{Diffusion {Measurements} in {Binary} {Liquid}
  {Mixtures} by {Raman} {Spectroscopy}}.
\newblock \bibinfo{journal}{Applied Spectroscopy} \bibinfo{volume}{61},
  \bibinfo{pages}{367--373}.
\newblock \URLprefix
  \url{http://journals.sagepub.com/doi/10.1366/000370207780466316},
  \DOIprefix\doi{10.1366/000370207780466316}.
\bibitem[{Carman(1967)}]{carman1967selfdiffusion}
\bibinfo{author}{Carman, P.C.}, \bibinfo{year}{1967}.
\newblock \bibinfo{title}{Self-diffusion and interdiffusion in complex-forming
  binary systems}.
\newblock \bibinfo{journal}{The Journal of Physical Chemistry}
  \bibinfo{volume}{71}, \bibinfo{pages}{2565--2572}.
\newblock \URLprefix \url{https://pubs.acs.org/doi/abs/10.1021/j100867a027},
  \DOIprefix\doi{10.1021/j100867a027}.
\bibitem[{Caticha(2007)}]{caticha2007information}
\bibinfo{author}{Caticha, A.}, \bibinfo{year}{2007}.
\newblock \bibinfo{title}{Information and {Entropy}}.
\newblock \bibinfo{journal}{AIP Conference Proceedings} \bibinfo{volume}{954},
  \bibinfo{pages}{11--22}.
\newblock \URLprefix \url{http://arxiv.org/abs/0710.1068},
  \DOIprefix\doi{10.1063/1.2821253}. \bibinfo{note}{arXiv: 0710.1068}.
\bibitem[{Cullinan and Toor(1965)}]{cullinan1965diffusion}
\bibinfo{author}{Cullinan, H.T.}, \bibinfo{author}{Toor, H.L.},
  \bibinfo{year}{1965}.
\newblock \bibinfo{title}{Diffusion in the {Three}-{Component} {Liquid}
  {System} {Acetone}-{Benzene}-{Carbon} {Tetrachloride}}.
\newblock \bibinfo{journal}{The Journal of Physical Chemistry}
  \bibinfo{volume}{69}, \bibinfo{pages}{3941--3949}.
\newblock \URLprefix \url{https://pubs.acs.org/doi/abs/10.1021/j100895a050},
  \DOIprefix\doi{10.1021/j100895a050}.
\bibitem[{D'Agostino et~al.(2011)D'Agostino, Mantle, Gladden and
  Moggridge}]{dagostino2011prediction}
\bibinfo{author}{D'Agostino, C.}, \bibinfo{author}{Mantle, M.D.},
  \bibinfo{author}{Gladden, L.F.}, \bibinfo{author}{Moggridge, G.D.},
  \bibinfo{year}{2011}.
\newblock \bibinfo{title}{Prediction of binary diffusion coefficients in
  non-ideal mixtures from {NMR} data: {Hexane}{\textendash}nitrobenzene near
  its consolute point}.
\newblock \bibinfo{journal}{Chemical Engineering Science} \bibinfo{volume}{66},
  \bibinfo{pages}{3898--3906}.
\newblock \URLprefix
  \url{https://www.sciencedirect.com/science/article/pii/S0009250911003150},
  \DOIprefix\doi{https://doi.org/10.1016/j.ces.2011.05.014}.
\bibitem[{D'Agostino et~al.(2013)D'Agostino, Stephens, Parkinson, Mantle,
  Gladden and Moggridge}]{dagostino2013prediction}
\bibinfo{author}{D'Agostino, C.}, \bibinfo{author}{Stephens, J.},
  \bibinfo{author}{Parkinson, J.}, \bibinfo{author}{Mantle, M.},
  \bibinfo{author}{Gladden, L.}, \bibinfo{author}{Moggridge, G.},
  \bibinfo{year}{2013}.
\newblock \bibinfo{title}{Prediction of the mutual diffusivity in
  acetone{\textendash}chloroform liquid mixtures from the tracer diffusion
  coefficients}.
\newblock \bibinfo{journal}{Chemical Engineering Science} \bibinfo{volume}{95},
  \bibinfo{pages}{43--47}.
\newblock \URLprefix
  \url{https://linkinghub.elsevier.com/retrieve/pii/S0009250913002157},
  \DOIprefix\doi{10.1016/j.ces.2013.03.033}.
\bibitem[{Darken(1948)}]{darken1948diffusion}
\bibinfo{author}{Darken, L.S.}, \bibinfo{year}{1948}.
\newblock \bibinfo{title}{Diffusion, mobility and their interrelation through
  free energy in binary metallic systems}.
\newblock \bibinfo{journal}{Trans. Aime} \bibinfo{volume}{175},
  \bibinfo{pages}{184--201}.
\bibitem[{Dhatt and Touzot(1984)}]{dhatt1984thefinite}
\bibinfo{author}{Dhatt, G.}, \bibinfo{author}{Touzot, G.},
  \bibinfo{year}{1984}.
\newblock \bibinfo{title}{The {Finite} {Element} {Method} {Displayed}}.
\newblock \bibinfo{publisher}{Wiley}.
\bibitem[{Einstein et~al.(1956)Einstein, F{\"u}rth and
  Cowper}]{einstein1956investigations}
\bibinfo{author}{Einstein, A.}, \bibinfo{author}{F{\"u}rth, R.},
  \bibinfo{author}{Cowper, A.D.}, \bibinfo{year}{1956}.
\newblock \bibinfo{title}{Investigations on the theory of brownian movement}.
\newblock \bibinfo{publisher}{Dover}, \bibinfo{address}{New York}.
\bibitem[{Fick(1995)}]{fick1995onliquid}
\bibinfo{author}{Fick, A.}, \bibinfo{year}{1995}.
\newblock \bibinfo{title}{On liquid diffusion}.
\newblock \bibinfo{journal}{Journal of Membrane Science} \bibinfo{volume}{100},
  \bibinfo{pages}{33--38}.
\newblock \URLprefix
  \url{https://linkinghub.elsevier.com/retrieve/pii/037673889400230V},
  \DOIprefix\doi{10.1016/0376-7388(94)00230-V}.
\bibitem[{Gmehling et~al.(1993)Gmehling, Onken and
  Arlt}]{gmehling1993vapourliquid}
\bibinfo{author}{Gmehling, J.}, \bibinfo{author}{Onken, U.},
  \bibinfo{author}{Arlt, W.}, \bibinfo{year}{1993}.
\newblock \bibinfo{title}{Vapour-liquid equilibrium data collection}.
\newblock \bibinfo{publisher}{Dechema}.
\bibitem[{Guevara-Carrion et~al.(2016)Guevara-Carrion, Janzen,
  Mu{\~n}oz-Mu{\~n}oz and Vrabec}]{guevara-carrion2016mutualdiffusion}
\bibinfo{author}{Guevara-Carrion, G.}, \bibinfo{author}{Janzen, T.},
  \bibinfo{author}{Mu{\~n}oz-Mu{\~n}oz, Y.M.}, \bibinfo{author}{Vrabec, J.},
  \bibinfo{year}{2016}.
\newblock \bibinfo{title}{Mutual diffusion of binary liquid mixtures containing
  methanol, ethanol, acetone, benzene, cyclohexane, toluene, and carbon
  tetrachloride}.
\newblock \bibinfo{journal}{The Journal of Chemical Physics}
  \bibinfo{volume}{144}, \bibinfo{pages}{124501}.
\newblock \URLprefix \url{http://aip.scitation.org/doi/10.1063/1.4943395},
  \DOIprefix\doi{10.1063/1.4943395}.
\bibitem[{Haase(1990)}]{haase1990thermodynamics2}
\bibinfo{author}{Haase, R.R.}, \bibinfo{year}{1990}.
\newblock \bibinfo{title}{Thermodynamics of Irreversible Processes}.
\newblock \bibinfo{publisher}{Dover}.
\bibitem[{Harris et~al.(1970)Harris, Pua and Dunlop}]{harris1970mutualand}
\bibinfo{author}{Harris, K.R.}, \bibinfo{author}{Pua, C.K.N.},
  \bibinfo{author}{Dunlop, P.J.}, \bibinfo{year}{1970}.
\newblock \bibinfo{title}{Mutual and tracer diffusion coefficients and
  frictional coefficients for the systems benzene-chlorobenzene,
  benzene-n-hexane, and benzene-n-heptane at 25.deg.}
\newblock \bibinfo{journal}{The Journal of Physical Chemistry}
  \bibinfo{volume}{74}, \bibinfo{pages}{3518--3529}.
\newblock \URLprefix \url{https://pubs.acs.org/doi/abs/10.1021/j100713a015},
  \DOIprefix\doi{10.1021/j100713a015}.
\bibitem[{Haynes et~al.(2016)Haynes, Lide and Bruno}]{haynes2016crchandbook}
\bibinfo{editor}{Haynes, W.M.}, \bibinfo{editor}{Lide, D.R.},
  \bibinfo{editor}{Bruno, T.J.} (Eds.), \bibinfo{year}{2016}.
\newblock \bibinfo{title}{{CRC} {Handbook} of {Chemistry} and {Physics}}.
\newblock \bibinfo{edition}{97th} ed., \bibinfo{publisher}{CRC Press}.
\bibitem[{Howie(2002)}]{howie2002interpreting}
\bibinfo{author}{Howie, D.}, \bibinfo{year}{2002}.
\newblock \bibinfo{title}{Interpreting probability: controversies and
  developments in the early twentieth century}.
\newblock Cambridge studies in probability, induction, and decision theory,
  \bibinfo{publisher}{Cambridge University Press}, \bibinfo{address}{Cambridge
  New York}.
\bibitem[{Hsu and Chen(1998)}]{hsu1998correlation}
\bibinfo{author}{Hsu, Y.D.}, \bibinfo{author}{Chen, Y.P.},
  \bibinfo{year}{1998}.
\newblock \bibinfo{title}{Correlation of the mutual diffusion coefficients of
  binary liquid mixtures}.
\newblock \bibinfo{journal}{Fluid Phase Equilibria} \bibinfo{volume}{152},
  \bibinfo{pages}{149--168}.
\newblock \URLprefix
  \url{https://linkinghub.elsevier.com/retrieve/pii/S0378381298003756},
  \DOIprefix\doi{10.1016/S0378-3812(98)00375-6}.
\bibitem[{Jaynes(1989)}]{jaynes1989clearing}
\bibinfo{author}{Jaynes, E.T.}, \bibinfo{year}{1989}.
\newblock \bibinfo{title}{Clearing up mysteries{\textemdash}the original goal},
  in: \bibinfo{booktitle}{Maximum {Entropy} and {Bayesian} {Methods}:
  {Cambridge}, {England}, 1988}. \bibinfo{publisher}{Springer}, pp.
  \bibinfo{pages}{1--27}.
\bibitem[{Jaynes and Bretthorst(2003)}]{jaynes2003probability}
\bibinfo{author}{Jaynes, E.T.}, \bibinfo{author}{Bretthorst, G.L.},
  \bibinfo{year}{2003}.
\newblock \bibinfo{title}{Probability theory: the logic of science}.
\newblock \bibinfo{publisher}{Cambridge university press},
  \bibinfo{address}{Cambridge}.
\bibitem[{Jianmin et~al.(2008)Jianmin, Shenglong and
  Xianjin}]{jianmin2008phenomenological2}
\bibinfo{author}{Jianmin, Y.}, \bibinfo{author}{Shenglong, L.},
  \bibinfo{author}{Xianjin, L.}, \bibinfo{year}{2008}.
\newblock \bibinfo{title}{Phenomenological {Models} of {Diffusivities} {Based}
  on {Local} {Composition}}.
\newblock \bibinfo{journal}{Advances in Natural Science} \bibinfo{volume}{1},
  \bibinfo{pages}{24--38}.
\newblock \URLprefix
  \url{http://www.cscanada.net/index.php/ans/article/view/j.ans.1715787020080101.004.004/36}.
\bibitem[{Kamei and Oishi(1970)}]{kamei1970selfdiffusion}
\bibinfo{author}{Kamei, Y.}, \bibinfo{author}{Oishi, Y.}, \bibinfo{year}{1970}.
\newblock \bibinfo{title}{{Self}-{Diffusion} {Coefficients} {of} {Water} {and}
  {Acetone} {in} {Water}-{Acetone} {System}}.
\newblock \bibinfo{note}{Issue: 4 Pages: 403 Publication Title: Nippon Kagaku
  Zasshi Volume: 91}.
\bibitem[{Kontogeorgis and Folas(2010)}]{kontogeorgis2010activity}
\bibinfo{author}{Kontogeorgis, G.}, \bibinfo{author}{Folas, G.},
  \bibinfo{year}{2010}.
\newblock \bibinfo{title}{Activity {Coefficient} {Models} {Part} 2: {Local}
  {Composition} {Models}, from {Wilson} and {NRTL} to {UNIQUAC} and {UNIFAC}},
  in: \bibinfo{booktitle}{Thermodynamic {Models} for {Industrial}
  {Applications}}. \bibinfo{publisher}{John Wiley \& Sons, Ltd}, pp.
  \bibinfo{pages}{109--157}.
\newblock \URLprefix
  \url{https://onlinelibrary.wiley.com/doi/abs/10.1002/9780470747537.ch5},
  \DOIprefix\doi{https://doi.org/10.1002/9780470747537.ch5}.
\bibitem[{Krishna(2015)}]{krishna2015uphilldiffusion}
\bibinfo{author}{Krishna, R.}, \bibinfo{year}{2015}.
\newblock \bibinfo{title}{Uphill diffusion in multicomponent mixtures}.
\newblock \bibinfo{journal}{Chemical Society Reviews} \bibinfo{volume}{44},
  \bibinfo{pages}{2812--2836}.
\newblock \URLprefix \url{http://xlink.rsc.org/?DOI=C4CS00440J},
  \DOIprefix\doi{10.1039/C4CS00440J}.
\bibitem[{Kutsyk et~al.(2021)Kutsyk, Ilchenko, Nikonova and
  Obukhovsky}]{kutsyk2021mixingdynamics2}
\bibinfo{author}{Kutsyk, A.M.}, \bibinfo{author}{Ilchenko, O.O.},
  \bibinfo{author}{Nikonova, V.V.}, \bibinfo{author}{Obukhovsky, V.V.},
  \bibinfo{year}{2021}.
\newblock \bibinfo{title}{Mixing dynamics of diethyl ether and chloroform}.
\newblock \bibinfo{journal}{Journal of Molecular Liquids}
  \bibinfo{volume}{339}, \bibinfo{pages}{116687}.
\newblock \URLprefix
  \url{https://www.sciencedirect.com/science/article/pii/S0167732221014112},
  \DOIprefix\doi{https://doi.org/10.1016/j.molliq.2021.116687}.
\bibitem[{Li et~al.(2001)Li, Liu and Hu}]{li2001amutualdiffusioncoefficient}
\bibinfo{author}{Li, J.}, \bibinfo{author}{Liu, H.}, \bibinfo{author}{Hu, Y.},
  \bibinfo{year}{2001}.
\newblock \bibinfo{title}{A mutual-diffusion-coefficient model based on local
  composition}.
\newblock \bibinfo{journal}{Fluid Phase Equilibria} \bibinfo{volume}{187-188},
  \bibinfo{pages}{193--208}.
\newblock \URLprefix
  \url{https://linkinghub.elsevier.com/retrieve/pii/S0378381201005350},
  \DOIprefix\doi{10.1016/S0378-3812(01)00535-0}.
\bibitem[{McCall and Anderson(1966)}]{mccall1966selfdiffusion}
\bibinfo{author}{McCall, D.W.}, \bibinfo{author}{Anderson, E.W.},
  \bibinfo{year}{1966}.
\newblock \bibinfo{title}{Self-{Diffusion} in {Cyclohexane}-{Benzene}
  {Solutions}}.
\newblock \bibinfo{journal}{The Journal of Physical Chemistry}
  \bibinfo{volume}{70}, \bibinfo{pages}{601--602}.
\newblock \URLprefix \url{http://pubs.acs.org/doi/abs/10.1021/j100874a514},
  \DOIprefix\doi{10.1021/j100874a514}.
\bibitem[{Mills(1965)}]{mills1965theintradiffusion}
\bibinfo{author}{Mills, R.}, \bibinfo{year}{1965}.
\newblock \bibinfo{title}{The {Intradiffusion} $^{\textrm{1,2}}$ and {Derived}
  {Frictional} {Coefficients} for {Benzene} and {Cyclohexane} in {Their}
  {Mixtures} at 25{\textdegree}}.
\newblock \bibinfo{journal}{The Journal of Physical Chemistry}
  \bibinfo{volume}{69}, \bibinfo{pages}{3116--3119}.
\newblock \URLprefix \url{https://pubs.acs.org/doi/abs/10.1021/j100893a051},
  \DOIprefix\doi{10.1021/j100893a051}.
\bibitem[{Mills and Hertz(1980)}]{mills1980application}
\bibinfo{author}{Mills, R.}, \bibinfo{author}{Hertz, H.G.},
  \bibinfo{year}{1980}.
\newblock \bibinfo{title}{Application of the velocity cross-correlation method
  to binary nonelectrolyte mixtures}.
\newblock \bibinfo{journal}{The Journal of Physical Chemistry}
  \bibinfo{volume}{84}, \bibinfo{pages}{220--224}.
\newblock \URLprefix \url{https://api.semanticscholar.org/CorpusID:95866746}.
\bibitem[{Moggridge(2012a)}]{moggridge2012prediction}
\bibinfo{author}{Moggridge, G.}, \bibinfo{year}{2012}a.
\newblock \bibinfo{title}{Prediction of the mutual diffusivity in binary liquid
  mixtures containing one dimerising species, from the tracer diffusion
  coefficients}.
\newblock \bibinfo{journal}{Chemical Engineering Science} \bibinfo{volume}{76},
  \bibinfo{pages}{199--205}.
\newblock \URLprefix
  \url{https://linkinghub.elsevier.com/retrieve/pii/S0009250912002400},
  \DOIprefix\doi{10.1016/j.ces.2012.04.014}.
\bibitem[{Moggridge(2012b)}]{moggridge2012prediction2}
\bibinfo{author}{Moggridge, G.}, \bibinfo{year}{2012}b.
\newblock \bibinfo{title}{Prediction of the mutual diffusivity in binary
  non-ideal liquid mixtures from the tracer diffusion coefficients}.
\newblock \bibinfo{journal}{Chemical Engineering Science} \bibinfo{volume}{71},
  \bibinfo{pages}{226--238}.
\newblock \URLprefix
  \url{https://linkinghub.elsevier.com/retrieve/pii/S0009250911008724},
  \DOIprefix\doi{10.1016/j.ces.2011.12.016}.
\bibitem[{Monakhova et~al.(2014)Monakhova, Pozharov, Zakharova, Khvorostova,
  Markin, Lachenmeier, Kuballa and
  Mushtakova}]{monakhova2014associationhydrogen}
\bibinfo{author}{Monakhova, Y.B.}, \bibinfo{author}{Pozharov, M.V.},
  \bibinfo{author}{Zakharova, T.V.}, \bibinfo{author}{Khvorostova, E.K.},
  \bibinfo{author}{Markin, A.V.}, \bibinfo{author}{Lachenmeier, D.W.},
  \bibinfo{author}{Kuballa, T.}, \bibinfo{author}{Mushtakova, S.P.},
  \bibinfo{year}{2014}.
\newblock \bibinfo{title}{Association/{Hydrogen} {Bonding} of {Acetone} in
  {Polar} and {Non}-polar {Solvents}: {NMR} and {NIR} {Spectroscopic}
  {Investigations} with {Chemometrics}}.
\newblock \bibinfo{journal}{Journal of Solution Chemistry}
  \bibinfo{volume}{43}, \bibinfo{pages}{1963--1980}.
\newblock \URLprefix \url{http://link.springer.com/10.1007/s10953-014-0249-1},
  \DOIprefix\doi{10.1007/s10953-014-0249-1}.
\bibitem[{Nauman and Savoca(2001)}]{nauman2001anengineering}
\bibinfo{author}{Nauman, E.B.}, \bibinfo{author}{Savoca, J.},
  \bibinfo{year}{2001}.
\newblock \bibinfo{title}{An engineering approach to an unsolved problem in
  multicomponent diffusion}.
\newblock \bibinfo{journal}{AIChE Journal} \bibinfo{volume}{47},
  \bibinfo{pages}{1016--1021}.
\newblock \URLprefix \url{http://doi.wiley.com/10.1002/aic.690470508},
  \DOIprefix\doi{10.1002/aic.690470508}.
\bibitem[{Obukhovsky et~al.(2017)Obukhovsky, Kutsyk, Nikonova and
  Ilchenko}]{obukhovsky2017nonlinear}
\bibinfo{author}{Obukhovsky, V.V.}, \bibinfo{author}{Kutsyk, A.M.},
  \bibinfo{author}{Nikonova, V.V.}, \bibinfo{author}{Ilchenko, O.O.},
  \bibinfo{year}{2017}.
\newblock \bibinfo{title}{Nonlinear diffusion in multicomponent liquid
  solutions}.
\newblock \bibinfo{journal}{Physical Review E} \bibinfo{volume}{95}.
\newblock \URLprefix \url{https://link.aps.org/doi/10.1103/PhysRevE.95.022133},
  \DOIprefix\doi{10.1103/PhysRevE.95.022133}.
\bibitem[{Po{\v z}ar et~al.(2015)Po{\v z}ar, Seguier, Guerche, Mazighi,
  Zorani{\'c}, Mijakovi{\'c}, Ke{\v z}i{\'c}-Lovrin{\v c}evi{\'c}, Sokoli{\'c}
  and Perera}]{povzar2015simpleand}
\bibinfo{author}{Po{\v z}ar, M.}, \bibinfo{author}{Seguier, J.B.},
  \bibinfo{author}{Guerche, J.}, \bibinfo{author}{Mazighi, R.},
  \bibinfo{author}{Zorani{\'c}, L.}, \bibinfo{author}{Mijakovi{\'c}, M.},
  \bibinfo{author}{Ke{\v z}i{\'c}-Lovrin{\v c}evi{\'c}, B.},
  \bibinfo{author}{Sokoli{\'c}, F.}, \bibinfo{author}{Perera, A.},
  \bibinfo{year}{2015}.
\newblock \bibinfo{title}{Simple and complex disorder in binary mixtures with
  benzene as a common solvent}.
\newblock \bibinfo{journal}{Physical Chemistry Chemical Physics}
  \bibinfo{volume}{17}, \bibinfo{pages}{9885--9898}.
\newblock \URLprefix \url{http://xlink.rsc.org/?DOI=C4CP05970K},
  \DOIprefix\doi{10.1039/C4CP05970K}.
\bibitem[{Price and Romdhane(2003)}]{price2003multicomponent}
\bibinfo{author}{Price, P.E.}, \bibinfo{author}{Romdhane, I.H.},
  \bibinfo{year}{2003}.
\newblock \bibinfo{title}{Multicomponent diffusion theory and its applications
  to polymer-solvent systems}.
\newblock \bibinfo{journal}{AIChE Journal} \bibinfo{volume}{49},
  \bibinfo{pages}{309--322}.
\newblock \URLprefix \url{http://doi.wiley.com/10.1002/aic.690490204},
  \DOIprefix\doi{10.1002/aic.690490204}.
\bibitem[{Renon and Prausnitz(1968)}]{renon1968localcompositions}
\bibinfo{author}{Renon, H.}, \bibinfo{author}{Prausnitz, J.M.},
  \bibinfo{year}{1968}.
\newblock \bibinfo{title}{Local compositions in thermodynamic excess functions
  for liquid mixtures}.
\newblock \bibinfo{journal}{AIChE Journal} \bibinfo{volume}{14},
  \bibinfo{pages}{135--144}.
\newblock \URLprefix
  \url{https://aiche.onlinelibrary.wiley.com/doi/abs/10.1002/aic.690140124},
  \DOIprefix\doi{https://doi.org/10.1002/aic.690140124}.
  \bibinfo{note}{\_eprint:
  https://aiche.onlinelibrary.wiley.com/doi/pdf/10.1002/aic.690140124}.
\bibitem[{Rodwin et~al.(1965)Rodwin, Harpst and Lyons}]{rodwin1965diffusion}
\bibinfo{author}{Rodwin, L.}, \bibinfo{author}{Harpst, J.A.},
  \bibinfo{author}{Lyons, P.A.}, \bibinfo{year}{1965}.
\newblock \bibinfo{title}{Diffusion in the {System}
  {Cyclohexane}{\textemdash}{Benzene}}.
\newblock \bibinfo{journal}{The Journal of Physical Chemistry}
  \bibinfo{volume}{69}, \bibinfo{pages}{2783--2785}.
\newblock \URLprefix \url{http://pubs.acs.org/doi/abs/10.1021/j100892a503},
  \DOIprefix\doi{10.1021/j100892a503}.
\bibitem[{Sanni et~al.(1971)Sanni, Fell and Hutchison}]{sanni1971diffusion}
\bibinfo{author}{Sanni, S.A.}, \bibinfo{author}{Fell, C.J.D.},
  \bibinfo{author}{Hutchison, H.P.}, \bibinfo{year}{1971}.
\newblock \bibinfo{title}{Diffusion coefficients and densities for binary
  organic liquid mixtures}.
\newblock \bibinfo{journal}{Journal of Chemical \& Engineering Data}
  \bibinfo{volume}{16}, \bibinfo{pages}{424--427}.
\newblock \URLprefix \url{https://pubs.acs.org/doi/abs/10.1021/je60051a009},
  \DOIprefix\doi{10.1021/je60051a009}.
\bibitem[{Sanni and Hutchison(1973)}]{sanni1973diffusivities}
\bibinfo{author}{Sanni, S.A.}, \bibinfo{author}{Hutchison, P.},
  \bibinfo{year}{1973}.
\newblock \bibinfo{title}{Diffusivities and densities for binary liquid
  mixtures}.
\newblock \bibinfo{journal}{Journal of Chemical \& Engineering Data}
  \bibinfo{volume}{18}, \bibinfo{pages}{317--322}.
\newblock \URLprefix \url{http://pubs.acs.org/doi/abs/10.1021/je60058a028},
  \DOIprefix\doi{10.1021/je60058a028}.
\bibitem[{Speagle(2020)}]{speagle2020dynesty}
\bibinfo{author}{Speagle, J.S.}, \bibinfo{year}{2020}.
\newblock \bibinfo{title}{dynesty: {A} {Dynamic} {Nested} {Sampling} {Package}
  for {Estimating} {Bayesian} {Posteriors} and {Evidences}}.
\newblock \bibinfo{journal}{Monthly Notices of the Royal Astronomical Society}
  \bibinfo{volume}{493}, \bibinfo{pages}{3132--3158}.
\newblock \URLprefix \url{http://arxiv.org/abs/1904.02180},
  \DOIprefix\doi{10.1093/mnras/staa278}. \bibinfo{note}{arXiv: 1904.02180}.
\bibitem[{Taylor and Krishna(1993)}]{taylor1993multicomponent}
\bibinfo{author}{Taylor, R.}, \bibinfo{author}{Krishna, R.},
  \bibinfo{year}{1993}.
\newblock \bibinfo{title}{Multicomponent mass transfer}.
\newblock Wiley series in chemical engineering, \bibinfo{publisher}{Wiley},
  \bibinfo{address}{New York}.
\bibitem[{Tomza et~al.(2019)Tomza, Wrzeszcz and Czarnecki}]{tomza2019tracking}
\bibinfo{author}{Tomza, P.}, \bibinfo{author}{Wrzeszcz, W.},
  \bibinfo{author}{Czarnecki, M.A.}, \bibinfo{year}{2019}.
\newblock \bibinfo{title}{Tracking small heterogeneity in binary mixtures of
  aliphatic and aromatic hydrocarbons: {NIR} spectroscopic, {2DCOS} and
  {MCR}-{ALS} studies}.
\newblock \bibinfo{journal}{Journal of Molecular Liquids}
  \bibinfo{volume}{276}, \bibinfo{pages}{947--953}.
\newblock \URLprefix
  \url{https://www.sciencedirect.com/science/article/pii/S0167732218348827},
  \DOIprefix\doi{https://doi.org/10.1016/j.molliq.2018.12.131}.
\bibitem[{Tyn and Calus(1975)}]{tyn1975temperature}
\bibinfo{author}{Tyn, M.T.}, \bibinfo{author}{Calus, W.F.},
  \bibinfo{year}{1975}.
\newblock \bibinfo{title}{Temperature and concentration dependence of mutual
  diffusion coefficients of some binary liquid systems}.
\newblock \bibinfo{journal}{Journal of Chemical \& Engineering Data}
  \bibinfo{volume}{20}, \bibinfo{pages}{310--316}.
\newblock \URLprefix \url{https://pubs.acs.org/doi/abs/10.1021/je60066a009},
  \DOIprefix\doi{10.1021/je60066a009}.
\bibitem[{Vignes(1966)}]{vignes1966diffusion}
\bibinfo{author}{Vignes, A.}, \bibinfo{year}{1966}.
\newblock \bibinfo{title}{Diffusion in {Binary} {Solutions}. {Variation} of
  {Diffusion} {Coefficient} with {Composition}}.
\newblock \bibinfo{journal}{Industrial \& Engineering Chemistry Fundamentals}
  \bibinfo{volume}{5}, \bibinfo{pages}{189--199}.
\newblock \URLprefix \url{https://pubs.acs.org/doi/abs/10.1021/i160018a007},
  \DOIprefix\doi{10.1021/i160018a007}.
\bibitem[{Vrentas and Vrentas(2007)}]{vrentas2007restrictions}
\bibinfo{author}{Vrentas, J.S.}, \bibinfo{author}{Vrentas, C.M.},
  \bibinfo{year}{2007}.
\newblock \bibinfo{title}{Restrictions on {Friction} {Coefficients} for
  {Binary} and {Ternary} {Diffusion}}.
\newblock \bibinfo{journal}{Industrial \& Engineering Chemistry Research}
  \bibinfo{volume}{46}, \bibinfo{pages}{3422--3428}.
\newblock \URLprefix \url{https://pubs.acs.org/doi/10.1021/ie061593a},
  \DOIprefix\doi{10.1021/ie061593a}.
\bibitem[{Weing{\"a}rtner(1990)}]{weingartner1990themicroscopic}
\bibinfo{author}{Weing{\"a}rtner, H.}, \bibinfo{year}{1990}.
\newblock \bibinfo{title}{The {Microscopic} {Basis} of {Self} {Diffusion} -
  {Mutual} {Diffusion} {Relationships} in {Binary} {Liquid} {Mixtures}}.
\newblock \bibinfo{journal}{Berichte der Bunsengesellschaft f{\"u}r
  physikalische Chemie} \bibinfo{volume}{94}, \bibinfo{pages}{358--364}.
\newblock \URLprefix \url{http://doi.wiley.com/10.1002/bbpc.19900940331},
  \DOIprefix\doi{10.1002/bbpc.19900940331}.
\bibitem[{Wilson(1964)}]{wilson1964vaporliquid}
\bibinfo{author}{Wilson, G.M.}, \bibinfo{year}{1964}.
\newblock \bibinfo{title}{Vapor-{Liquid} {Equilibrium}. {XI}. {A} {New}
  {Expression} for the {Excess} {Free} {Energy} of {Mixing}}.
\newblock \bibinfo{journal}{Journal of the American Chemical Society}
  \bibinfo{volume}{86}, \bibinfo{pages}{127--130}.
\newblock \URLprefix \url{http://pubs.acs.org/doi/abs/10.1021/ja01056a002},
  \DOIprefix\doi{10.1021/ja01056a002}.
\bibitem[{Zhou et~al.(2013)Zhou, Yuan, Zhang and Yu}]{zhou2013localcomposition}
\bibinfo{author}{Zhou, M.}, \bibinfo{author}{Yuan, X.}, \bibinfo{author}{Zhang,
  Y.}, \bibinfo{author}{Yu, K.T.}, \bibinfo{year}{2013}.
\newblock \bibinfo{title}{Local {Composition} {Based}
  {Maxwell}{\textendash}{Stefan} {Diffusivity} {Model} for {Binary} {Liquid}
  {Systems}}.
\newblock \bibinfo{journal}{Industrial \& Engineering Chemistry Research}
  \bibinfo{volume}{52}, \bibinfo{pages}{10845--10852}.
\newblock \URLprefix \url{http://pubs.acs.org/doi/10.1021/ie4010157},
  \DOIprefix\doi{10.1021/ie4010157}.
\bibitem[{Zhu et~al.(2015)Zhu, Moggridge and
  D{\textquoteright}Agostino}]{zhu2015alocal}
\bibinfo{author}{Zhu, Q.}, \bibinfo{author}{Moggridge, G.D.},
  \bibinfo{author}{D{\textquoteright}Agostino, C.}, \bibinfo{year}{2015}.
\newblock \bibinfo{title}{A local composition model for the prediction of
  mutual diffusion coefficients in binary liquid mixtures from tracer diffusion
  coefficients}.
\newblock \bibinfo{journal}{Chemical Engineering Science}
  \bibinfo{volume}{132}, \bibinfo{pages}{250--258}.
\newblock \URLprefix
  \url{https://linkinghub.elsevier.com/retrieve/pii/S0009250915002821},
  \DOIprefix\doi{10.1016/j.ces.2015.04.021}.
\bibitem[{Zielinski and Hanley(1999)}]{zielinski1999practical}
\bibinfo{author}{Zielinski, J.M.}, \bibinfo{author}{Hanley, B.F.},
  \bibinfo{year}{1999}.
\newblock \bibinfo{title}{Practical friction-based approach to modeling
  multicomponent diffusion}.
\newblock \bibinfo{journal}{AIChE Journal} \bibinfo{volume}{45},
  \bibinfo{pages}{1--12}.
\newblock \URLprefix \url{http://doi.wiley.com/10.1002/aic.690450102},
  \DOIprefix\doi{10.1002/aic.690450102}.

\end{thebibliography}

\end{document}